\documentclass{biophys-new}
\usepackage[utf8]{inputenc}
\usepackage{graphicx}
\usepackage[colorlinks,allcolors=cyan!70!black]{hyperref}

\usepackage{lipsum}

\title{Chromatin and cytoskeletal tethering determine nuclear morphology in progerin expressing cells}
\runningtitle{Biophysical Journal Template} 

\author{M. C. Lionetti, S. Bonfanti, M. R. Fumagalli, Z. Budrikis, F. Font Clos,  G. Costantini, 
O. Chepizhko, S. Zapperi, C. A. M. La Porta}

\runningauthor{M. C. Lionetti et al.} 

\papertype{Article}

\begin{document}

\begin{frontmatter}

\begin{abstract}
The nuclear morphology of eukaryotic cells is determined by the interplay between the lamina forming the nuclear skeleton, the chromatin inside the nucleus and the coupling with the cytoskeleton. Nuclear alterations are often associated with pathological conditions as in the Hutchinson-Gilford progeria syndrome (HGPS) where a mutation in the lamin A gene yields an altered form of the protein, named progerin, and  an aberrant nuclear shape. Here, we introduce an inducible cellular model of HGPS in HeLa cells where increased progerin expression leads to alterations in the coupling of the lamin shell with cytoskeletal/chromatin tethers as well as with Polycomb-group (PcG) proteins. Furthermore, our experiments show that progerin expression leads to enhanced nuclear shape fluctuations in response to cytoskeletal activity. To interpret the experimental results, we introduce a computational model of the cell nucleus that includes explicitly chromatin fibers, the nuclear shell and the coupling with the cytoskeleton. The model allows us to investigate how the geometrical organization of chromatin-lamin tether affects nuclear morphology and shape fluctuations.  In sum, our findings highlight the crucial role played by lamin-chromatin and lamin-cytoskeletal alterations in determining nuclear shape morphology and in  affecting cellular functions and gene regulation. 
\end{abstract}

\begin{sigstatement}
Hutchinson-Gilford progeria syndrome (HGPS) is a rare disease characterized by accelerated aging  due to a mutation of the lamin A gene, leading to an aberrant protein --- named progerin --- and to the formation of nuclear blebs.  We combine experiments on a cellular model reproducing HGPS cells and numerical simulations of nuclear mechanics to study the biological and biophysical effects of progerin expression on nuclear morphology and functioning. Our results show that progerin induces key changes in the mechanical tethering between cytoskeleton, lamins and chromatin producing nuclear shape alterations and affecting gene regulation.
\end{sigstatement}
\end{frontmatter}

\section*{Introduction}

The cell nucleus is of crucial importance in eukaryotic cells since it harbors and preserves the genetic information encoded into chromatin fibers. The protective role of the nucleus is ensured by its structural and mechanical features, mainly due to a rigid nuclear shell composed by a dense network of filaments assembled from lamin proteins. The relative concentration of the different types of lamins (e. g. lamin A, B and C) composing the nuclear skeleton can have an important effect on  the elastic and visco-elastic response of the nucleus \cite{lammerding2004,lammerding2006}, affecting gene expression and cell fate too \cite{Swift2013}. There is increasing awareness that the mechanical properties of the nucleus are not only determined by the properties of the nuclear skeleton, but crucially depend on the tethering between chromatin and lamins \cite{Schreiner2015} and on the coupling between lamins and cytoskeleton \cite{Makhija2016,Chu2017}. In particular, lamin A is involved in a complex molecular interface between the inner membrane of the nuclear envelope and chromatin fibers \cite{goldman2004} and the integrity of this interface is crucial for correct  chromatin functioning during cell cycle and apoptosis. Lamin mutations can thus lead to severe morphological and mechanical nuclear alterations, often associated with severe pathological conditions \cite{zwerger2011}. For instance, the Hutchinson-Gilford progeria syndrome (HGPS) is a rare autosomal-dominant disease characterized by accelerated aging due to a {\it de novo} mutation in LMNA gene, leading to an aberrant form of the protein, known as progerin \cite{Eriksson2003}.  Cells from HGPS patients typically display altered nuclear shapes with bleb-like protrusions.

While the genetic origin of HGPS has been clarified \cite{Eriksson2003}, the precise mechanism leading to the changes in nuclear morphology is still unknown.  In addition to blebs, nuclei from HGPS patients also display gross reorganization of the lamin architecture into distinct domains \cite{dahl2006} and mechanical stiffness alterations \cite{dahl2006,verstraeten2008}.  Recent experiments using micropipette aspiration show that nuclei expressing exogenous progerin are also stiffer with respect to control cells \cite{booth2015}, resembling the HGPS phenotype\cite{dahl2006}. These works also found that inside the nucleus chromatin is softer and displays reduced mobility when progerin is expressed \cite{booth2015}. Interestingly, progerin expression is found to reduce the extent of mechanical force propagation to the nuclear interior from the cytoskeleton \cite{booth2015}. 

The mechanical coupling between cytoskeleton, nuclear envelope and chromatin fibers is 
mediated by a set of key proteins acting as connectors between the different elements. In particular, Sun proteins, located in the inner nuclear membrane, are known to interact with lamins in the nucleoplasm, while nesprin (KASH) recruits cytoskeletal components to the outer nuclear membrane.  Sun proteins also interact with the protein emerin \cite{Berk2013} which is located in the inner nuclear membrane and it is directly linked to lamin A \cite{Clements2000}. Emerin is known to act as a chromatin tether and its mutation  may cause Emery-Dreifuss Muscular Dystrophy (EDMD). It is interesting to note that EDMD may also display mutations  in nesprin-1 and -2 \cite{Zhang2007} and SUN1 \cite{Li2014}, leading to a softer nucleus and looser association with lamin A \cite{Markiewicz2002}.  Another key protein for nuclear mechanics is fascin, a F-actin-binding protein \cite{Vignjevic2006} which plays a role in the stabilization of filopodia \cite{Jayo2012} and in the molecular adhesion dynamics of migrating cells \cite{Elkhatib2014}. A recent paper shows that fascin can bind directly to nesprin-2 at the nuclear envelope and this interaction is independent of the role of fascin within filopodia at the cell periphery  \cite{Jayo2016}. Moreover, it has been shown that the disruption of this interaction induces important changes in nuclear shape and deformation \cite{Jayo2016}.  In particular, uncoupling the S39-phospho-fascin/nesprin-2 complex leads to a reduction of nuclear deformation and affects
in a significant way many functional properties, such as cell invasion \cite{Jayo2016}. 

Several computational models have been used to investigate nuclear mechanics, treating the nuclear skeleton either as a discretized continuous elastic shell \cite{Vaziri2007,Funkhouser2013} or  as a polymer network \cite{Wren2012,Banigan2017}.  Numerical simulations of a finite element model for the nuclear skeleton reveal that blebs are formed under the assumption that the shell is characterized by domains with different relative concentrations of lamin A and B \cite{Funkhouser2013}. According to this model, blebs would form in correspondence of domains that are rich in lamin A which would make them prone to expansion \cite{Funkhouser2013}. The model provides a good description of the morphology of the nuclei of lamin B silenced cells, but application to HGPS cells is not straightforward. Lamin domains have been observed in HGPS nuclei thanks to polarized light microscopy showing regions with distinct orientations in the lamin network \cite{dahl2006}. Different lamin domains in HGPS nuclei thus mostly differ due to their orientation rather than because of their relative lamin content, contrary to the model hypothesis \cite{Funkhouser2013}. The model did not consider the coupling between lamins and chromatin that is known to be important for nuclear morphology and mechanics \cite{Schreiner2015,Stephens2017,Stephens2018} 

In this paper, we investigated experimentally and computationally the role of chromatin and cytoskeletal tethering in affecting nuclear morphology and function in progerin-expressing cells. To tackle this question, we introduced an {\it in vitro} cellular model where progerin expression was induced at controlled and realistic levels in HeLa cells, thanks to the Tet-On technology. Our model was useful to overcome the limitations posed by the use of human fibroblasts obtained from HGPS patients that are only able to grow {\it in vitro} for few passages. We used our cellular model to investigate how the induction of progerin expression affect key proteins involved in the coupling of the nuclear shell with the cytoskeleton and with chromatin fibers. We then studied how alterations of these couplings impacted on critical nuclear functions and could possibly lead to changes in gene regulation. 

To interpret the experimental results, we constructed a mechanical model of the cell nucleus that included the coupling between nuclear lamina, chromatin fibers and cytoskeleton. To model chromatin fibers, we followed previous polymer models  \cite{Marko2008,Marti-Renom2011} that have been successfully used to study chromatin organization in the nucleus \cite{Dekker2013,Gibcus2018}. Chromatin fibers were tethered to a lamin shell, modeled by a triangulated surface endowed with stretching and bending rigidity, that is also elastically  tethered to an external set of oscillating points modeling, contractions of the cytoskeleton. The computational model allowed us to test {\it in silico} the effect of tether organization and strength on nuclear morphology and explain the presence of enhanced cytoskeleton-induced 
nuclear fluctuations that we observed experimentally when progerin expression was induced.  All together our findings highlight the important role played by chromatin and nuclear tethering in determining nuclear morphology and fluctuations with important implications for HGPS.

\section*{Materials and Methods}

\subsection*{Human Fibroblasts culture growth conditions}
Human fibroblasts from HGPS patient (HGSDNF167) and the healthy mother of this patient (HGMDF090) are obtained from the Progeria Foundation Cell and Tissue Bank. The HGPS patient was a male of 8 years old with the classical mutation  (heterozygous LMNA Exon 11 c1824 $C>T$ (p.Gly608Gly)).  The cells were maintained according to the protocol reported by  The Progeria Foundation: 15\% FBS, DMEM with 1\% antibiotics and 1\% L-glutamine. The cells were detached by trypsin  0.25\% EDTA and are maintained in culture for no more than five passages. 

\subsection*{Plasmids and Subcloning}
$\Delta$50-lamin A plasmid was obtained by Addgene (pEGFP-D50 lamin A, cod.17653). Plasmid expressing pTRE3G-mCherry vector was obtained by Clontech (cod. 631165). E. coli One Shot TOP10 bacteria (Invitrogen, C404006)  were used for transformation with pEGFP-$\Delta$50 lamin A and pTRE3G-mCherry vector.  Competent cells were expanded and selected in Luria Broth medium (Invitrogen, 12795-027) containing Kanamicin for pEGFP $\Delta$50-lamin A (Sigma Aldrich, K13747) and 100 $\mu$g/mL Ampicillin (Sigma Aldrich,  A5354 ) for pTRE3G-mCherry vector for 18h at $37^\circ$C.
AfeI/BamHI fragment containing the coding reading sequence of $\Delta$50 lamin A was excised by pEGFP-$\Delta$50 lamin A plasmid (AfeI cod.R0652S, New England Biolabs)  and subcloned into pTRE3G-mCherry vector linearised with EcorV (cod. R1095S, New England Biolabs) and BamHI (cod. R0136S, New England Biolabs) restriction enzymes according to the manufacturer's instructions.  DNA fragments and vectors were routinely analysed by electrophoresis on 1\% agarose gel in 1X TAE (Euroclone GellyPhor EMR010100). To recover DNA digested fragments and linearised vectors after electrophoresis and proceed with subcloning Low Melting was used (Euroclone Gellyphor cod. EMR911100). Promega T4 DNA ligase (cod. M1801, Promega) was used to allow ligation step in according to manufacturer's protocols. DNA fragments and vectors were routinely analysed by electrophoresis on 1\% agarose gel (Euroclone GellyPhor cod.EMR010100), in 1X TAE while Low Melting agarose gel (Euroclone Gellyphor cod.EMR911100) was used for the recovery of DNA fragments after electrophoresis. To purify DNA fragments from agarose gels QIAquick Gel Extraction Kit was used according to the manufacturer's instructions. To recover DNA digested fragments and linearised vectors after electrophoresis and proceed with subcloning,
Promega T4 DNA ligase (cod. M1801, Promega) was used to allow ligation step in according to manufacturer's protocols.
Promega PureYield Plasmid Miniprep and Midiprep (cod.A1330 and A2492, Promega) were used to purify plasmidic DNA.
The Product of this subcloning strategy was sequenced (Eurofins Genomics Service) using using IRES and SV40 primers 
(IRES2\_F: TGTGGAAAGAGTCAAATGGCT, SV40\_R : GGAACTGATGAATGGGAGCAG) to confirm that $\Delta$50-lamin A was cloned in-frame with the start codon at the IRES2/MCS junction (see Table~\ref{TableS1}).

\subsection*{Progerin inducible Tet-On HeLa cells}
HeLa 3G cells (Clontech cod. 631183 ) were cultured in DMEM (Euroclone cod. ECB7501L) supplemented with 10\% v/v Tet-free FBS (Euroclone, cod.ECS01821) 1\% Penincillin/Streptomycin and 1\% L-Glutamine at $37^\circ$C and 5\% $CO_2$ in humidified incubator immediately upon thawing without selective resistance.  To create a double stable inducible HeLa 3G cell line expressing $\Delta$50 lamin A, HeLa 3G cells were transfected with 5$\mu$g of pTRE3G-mCherry-$\Delta$50-lamin A using Xfect reagent (Clontech,cod. 631317) and 250 ng of puromycin linear selection marker (Clontech cod.631626) according to the manufacturer's instructions in order to generate a double-stable cell line expressing  both Tet-On 3G transactivator  and  higher levels of progerin in response to Doxycycline. After two weeks of drugs selection, 24 resistant colonies were picked up and screened for inducibility, with increasing doses of Doxycycline ( 0-1-5-10 and 20 ng/ml)  measuring mCherry and $\Delta$50 Lamin A by immunofluorescence and western blot. Two clones of the 24 screened were selected and used for further experiments. 48h post-transfection, the cells were splitted into 4 x 10 cm dishes and 0.5 $\mu$g/ml of Puromycin (Life Technologies cod. A11138-03 ) and 200$\mu$g/ml G418 (cod. A1720, Sigma) was added to select the positive clones. Drug-resistant colonies appeared 2 weeks after selection. Single clones were isolated using cloning cylinder (Sigma Catalog cod.C1059). When they reached confluence, the cells were split in 6-well plate for testing the expression of $\Delta$50 lamin A and for further maintenance. In order to express $\Delta$50-lamin A in dose dependent manner, the cells were induced with Doxycycline (Clontech cod.631311) at the concentration indicated in the figures and analysed 48h after the induction. HeLa 3G cells expressing $\Delta$50-lamin A were rountinely maintained in culture in DMEM (Euroclone cod. ECB7501L) supplemented with 10\% v/v Tet-free FBS (Euroclone cod.ECS01821), 200$\mu$/ml G418 and 0.25$\mu$g/ml Puromycin, 1\% Penincillin/Streptomycin and 1\% L-Glutamine at $37^\circ$C and 5\% $CO_2$.

\subsection*{$\Delta$50 lamin A and WT lamin A transient transfection}
Plasmid containing the human lamin A-C-18 (Addgene, cod. 55068) or progerin 
(pEGFP-$\Delta$50 lamin A, Addgene, cod.17653) were purchased by Addgene. E. coli One Shot TOP10 bacteria
(Invitrogen, cod.C404006) were transformed with plasmids, expanded and selected in Luria Broth medium 
(Invitrogen, cod. 12795-027) containing 50 $\mu$g/mL Kanamicin. 
PureYield Plasmid Miniprep and Midiprep (Promega, cod.A1330 and A2492) were used to purify plasmidic DNA according to the manufacturer's instructions.

HeLa cells were maintained in DMEM (Euroclone, cod. ECM0060L) medium with 10\% fetal bovine serum ( Euroclone, ECS0180L), 100 U/ml penicillin, 100 mg/ml streptomycin sulphate (Euroclone,cod. ECB3001D)
 and 2mM L-Glutamine ( Euroclone, cod. ECB3000D-20) at 37$^\circ$C in an atmosphere of 5\% CO2 and 95\% humidity.
 Cells seeded at an 70\% confluent onto 6-well plates were transiently transfected with pEGFP $\Delta$50lamin A (Addgene, cod.17653) or mEmerald-WT-lamin A (Addgene, cod.54139) using Xfect transfection reagent (Clontech, cod 631317). 
After 48 h from transfection cells were used for cytoskeleton pharmacological perturbation experiment.

\subsection*{Pharmacological perturbation of the cytosckeleton}
To study the role of cytoskeleton in this context, actin and myosin organization were perturbed by exposing pEGFP-$\Delta$50lamin A or mEmerald-WT-lamin A overexpressing cells to Blebbistatin, a myosin inhibitor, or to a cell-permeable inhibitor of formin-mediated actin nucleation and formin-mediated elongation of actin filaments, SMIFH2 \cite{Makhija2016}.
Subconfluent Hela cells expressing WT or $\Delta$50 lamin A were exposed to 25$\mu$M Blebbistatin (Sigma-Aldrich, B0560) for 30 min or to 20$\mu$M SMIFH2 (Sigma-Aldrich, S4826) for 1 hour at 37$^\circ$C and 5\% $CO_2$ in humidified incubator. 
Five minutes prior the end of the treatment, Hoechst (1:1000, Life technology, H3570) was added to the cell medium to counterstain nuclei. Immediately after the end of each treatment, 
 medium containing drugs was replaced with fresh medium and cells were time-lapse imaged (1 shoot every 15 minutes) for 1h using a Leica TCS NT confocal microscope (63X) with a z-stack of 0.5 $\mu$m.

\subsection*{Immunofluorescence} 
Subconfluent cells grown on glass coverslips were fixed with 3.7\% parafolmaldhehyde in PBS for 10min or with ice cold 100\% methanol for 5min at -20$^\circ$C, permeabilized with 0.5\% Triton X-100 in PBS for 5min at RT and incubated with 10\% goat serum in PBS for 1hr. The cells were stained with anti-lamin A (1:100 ab8980, Abcam) or anti-PanLamin (1: 50, ab20740, Abcam ) or anti-progerin (1:20 , ab66587, Abcam) or HP1 (1:250, ab109028, Abcam) overnight at $4^\circ$C.
Thus, after a brief washing with PBS, the cells are incubated with the secondary antibody (1:250, anti-rabbit,  ab150077, Abcam or anti-mouse, ab150113, Abcam) for 1 h. The nuclei are counterstained with DAPI and the slides mounted with Pro-long anti fade reagent (cod P36931, Life technologies). The images are acquired with a Leica TCS NT confocal microscope. Immunofluorescence intensity was estimated using customized ImageJ macro evaluating
single pixel fluorescence after subtracting the background noise. The average fluorescence was calculated on pixels that passed the background filtering. In all the analyzed frames, nuclei close to the edge of the image or superimposed were manually discarded.

\subsection*{Proximity Ligation Assay} 
Subconfluent cells were fixed on slides with ice cold 100\% methanol for 5 min at at -20$^\circ$C  and then incubated with Duolink Blocking Solution for 60 minutes at 37$^\circ$C in a humidity chamber.
Slides were then incubated  in a humidity chamber overnight at 4$^\circ$C with lamin A (1:100 cod.ab8980, Abcam),   PanLamin (1:50, ab20740, Abcam) antibody with  SUZ12 (1:800, mAb 3737, Cell Signaling) or BMI1 (1:600, mAb6964, Cell Signaling) or  SUN1 (1:200, ab103021, Abcam) or emerin (1:200 ab40688, Abcam).
After washing, samples were incubated in a pre-heated humidity chamber for 1 hour at 37$^\circ$C with anti-rabbit PLUS and anti-mouse MINUS PLA probes diluted 1:5.  PLA probes generates a fluorescent signal when bound to two different primary antibodies, raised in different species, that recognized two antigens in close proximity (less than 40nm). All the antibodies were diluted in Duolink Antibody Diluent. Ligation and amplification steps were performed according to manufacturer's instructions.  Slides were mounted with Duolink In Situ Mounting Medium with DAPI (DUO82040, Sigma-Aldrich). The images were acquired with a Leika TCS NT confocal microscope.

\subsection*{Proximity ligation assay analysis}
Immunofluorescence images were analyzed using existing and customized plugins of the bioimage informatics platform Icy (v.1.9.4 and 1.9.5~\cite{icy}) and custom python scripts. 
Reconstruction of progerin Tet-On HeLa nuclear envelope was performed on all the Z-stacks with Icy HK-Means and ActiveContour plugin using DAPI signal and the resulting 3D meshes were exported as VTK files.
A wide range of parameters was explored in order to ensure that our result were not affected by the specific parameters chosen for reconstruction.
Duolink spot recognition was performed separately on each nucleus with a semi-automatic protocol involving HK-means thresholding (ICY Thresholder plugin). The minimum size of each accepted spot was set to 70 px. Center of mass was used to determine the relative position of the spot respect to the nuclear envelope, and spots inside reconstructed nuclear mesh were taken into account.
In all the analyzed frames nuclei close to the borders or superimposed were manually discarded.

\subsection*{Immunoprecipitation assay}
Subconfluent cells were gently detached from the culture plates using a cell lifter in cold PBS,  collected in a 1.5 ml eppendorf and centrifuged at 3.5$\cdot 10^3$ x g for 5 min at 4$^\circ$C.  The pellet was resuspended in lysis buffer (2 mM EGTA ,0.5 mM EDTA, 0.5 mM PMSF, 1X TRITONX100, 1X Protease Cocktail Sigma Aldrich P8340) and incubated on ice for 30 min with periodic mix by vortex. 
The lysate was centrifuged at 14$\cdot 10^3$xg for 10 min at 4$^\circ$C, the supernatant transferred to a fresh tube and proteins concentration was measured by DC Protein Assay Kit (BioRad).  500 $\mu$g of total proteins was incubated with anti-fascin antibody (1:100,  ab126772, AbCam)  overnight at 4$^\circ$C under stirring.  50\% beads slurry of Protein A-Agarose (P9269, Sigma Aldrich)
was added to the lysate and reincubated with gentle rocking for 2 hours at  4$^\circ$C before centrifugation (14$\cdot 10^3$ x g for 
10 min at 4$^\circ$C). After three washes with 500$\mu$l of lysis buffer the sample was resuspended in 30$\mu$l 2X Laemmli sample buffer(2\% SDS, 20\% glycerol, and 125 mM Tris-HCl, pH 6.8). Samples were then heated to 90$^\circ$C for 5 minutes, and 15 $\mu$l per sample was loaded on 10\%SDS-PAGE for Western blot.
50 $\mu$g of the whole lysate for each sample were loaded on the same gel.

\subsection*{Western blot}
Subconfluent cells were lysed by boiling in a modified Laemmli sample buffer (2\% SDS, 20\% glycerol, and 125 mM Tris-HCl, pH6.8). The protein concentration was measured by DC Protein Assay Kit (BioRad). Equal amount of proteins were loaded on 10\% (or $7.5\%$ for IP assay) SDS-PAGE gel and transferred to a PVDF membrane (Trans-Blot Turbo Mini PVDF, BioRad).
After blocking with 5\% BSA/0.1 \% Tween20 in PBS for 1h at RT,the sheet was incubated with primary antibody overnight at 4$^\circ$C. The following antibodies and dilution were used in various experiments: 
anti-nesprin 2 (1:1000, cod. MABC86, Merck Millipore), anti-SUN1 (1:1000,Abcam ab103021),  anti-emerin (1:200 Abcam ab40688), anti S39-phospho-fascin antibody (1:10000, ab90648, AbCam),  anti-fascin (1:10000, AbCam, ab126772),  anti-SUZ12 (1:1000, cod. 3737, Cell Signaling), anti- Ezh2 (1:1000, cod. 5246 , Cell Signaling),  anti- Ring1A (1:1000, cod. 13069, Cell Signaling),  anti-RING1B (1:1000, cod. 5694, Cell Signaling), anti- BMI1 (1:1000, cod. 6964, Cell Signaling),  anti-vinculin (1:10000, cod. V9264, Sigma), anti-GAPDH(1:5000, cod. G9545, Sigma) or anti-$\beta$tubulin (1:5000, cod. T8328, Sigma) antibodies incubated for 1h at RT were used as housekeeping reference.

\subsection*{Wound healing assay}
For the migration assay, a wound is introduced in the central area of the confluent cell sheet by using a pipette tip and the migration followed by time-lapse imaging. The measurements of the velocity field were done using PIVlab app for Matlab \cite{PIV1}. The method is based on the comparison of the intensity fields of two consequent photographs of cells. The difference in the intensity is converted into velocity field measured in px/frames and then converted to $\mu$m/h \cite{Chepizhko2016}. The front position is quantified according to Ref. \cite{Chepizhko2016}

\subsection*{2D analysis of nuclear morphology}

\subsubsection*{Nuclear segmentation}
The transformation of fluorescence images into binary masks is done using a custom script in Matlab. Using a two-peak histogram of pixel intensity a threshold is defined for each image to cut away the background with uneven illumination. After applying this treshold, ``imbinarize'' function is used  with adaptive thresholding. Holes in the image are filled with ``imfill'' function. Little remaining noise is removed with ``bwareaopen'' function. The functions are a part of Matlab Image Processing Toolbox.
We process 96 fluoresence images automatically, and then manually check for segmentation errors. Usually cells that are undergoing division or cells in regions of lower fluorescence intensity are not well-segmented by our algorithm. After discarding segmentation errors by hand, we are left with binary masks of 297 No Doxy nuclei, 330 Doxy nuclei, 87 healthy nuclei, and 87 HGPS nuclei.

\subsubsection*{Computation of local curvature}
The curvature of a point in a planar curve $(x, y)$ is mathematically defined as

\begin{equation}
k = \frac{x' y'' - y' x'' }{\left( x'^2 + y'^2 \right)^{3/2}}
\label{eq:curvature}
\end{equation}
and corresponds to the inverse of the radius of a circle tangent to the point.
To estimate the curvature of nuclei along each point of their perimeter we first fit fourth-order splines to the border of the binary masks using ``UnivariateSpline'' function from SciPy python library \cite{SciPy} then evaluate expression (\ref{eq:curvature}) using the obtained differentiable curves.

\subsubsection*{Blebbiness index}
Negative curvature in 2-dimensional images is a signal of abnormal nuclear morphology, being 
associated with blebs and invaginations. Curvature changes along the perimeter of a nucleus, making it difficult to compare curvature estimates among large number of nuclei. To simplify the curvature values of nuclei along their perimeters into a single number for each nucleus, we define the blebbiness index $\phi$ as the weighted fraction of negative curvature,
\begin{equation}
\phi = \frac{\int_{\mathcal{C}_-} |k|}{\int_\mathcal{C} 1},
\end{equation}
where $\mathcal{C}$ is the perimeter of the nucleus and $\mathcal{C}_-$ is the region of the perimeter with negative curvature $k < 0$. The denominator $\int_\mathcal{C} 1$ is simply the length of the perimeter, and the numerator is the integral of the absolute value of the curvature along regions of negative curvature. In this way, regions of large negative curvature contribute more to the total than regions of negative but small curvature, making in turn the index $\phi$ robust to variability in the estimation of $k$. 

\subsubsection*{Automated bleb counting}
Automated bleb counting is a difficult problem. While the biological definition of bleb is well-stalished, a translation into quantitative terms admits different formulations. Here we estimate the number of blebs by counting the number of regions of negative local curvature $k$ along the nuclei perimeter. To avoid spurious results due to values of $k \simeq 0$, we define the number of blebs as the number of regions where the curvature is below a threshold $k_{\mathrm{min}} = -0.1$. Notice that we have a total of $N=801$ nuclei, making manual quantification unfeasible in this case.

\subsection*{3D analysis of nuclear morphology}

\subsubsection*{Mesh Reconstruction of the nuclear shell}
The Hela nucleus mesh reconstruction was performed using the bioimage informatics platform Icy \cite{icy}. 
Starting from the images with the nuclei and over-expressed lamin skeletons, we applied the Icy HK-Means plug-in to obtain   
three dimensional (3D) nuclear regions of interest (ROIs) for each temporal acquisition. 
This segmentation method uses, in fact, a K-Means classification to detect clustered objects corresponding, in our case, to the nucleus structures. 
The final 3D meshes of the outer nuclear membrane have been obtained thanks to the Icy Active Contours plug-in \cite{Dufour2011}.
The parameters used for the reconstruction are the default ones except for the values of contour smoothness, contour sampling and region sensitivity.
For these three parameters we used, respectively, the values in the ranges 0.028-0.032, 1.9-2.1 and 2-3.

\subsubsection*{Local displacements}
The total volume change $\Delta \mathrm{Vol}  = \mathrm{Vol}(M_{1}) -\mathrm{Vol}(M_{0})$ between two meshes $M_{0}$, $M_{1}$ of a given cell at different time-points is a global measure that gives only summarized information on morphological changes . While $\Delta \mathrm{Vol}$ can reveal key information in some situations, most local morphological changes like [XX, YY and ZZ (blebs etc)] can take place at fixed volume, remaining blind to $\Delta \mathrm{Vol}$. To circumvent this issue and detect local morphological changes, we develop the concept of local displacements $d_{j}$, which can be interpreted as the distance each face of $M_{0}$ should move to turn $M_{0}$ into $M_{1}$. 

Alternatively, one can think of local displacements as a decomposition of the total volume change,  
\begin{equation}
\sum_j d_j s_j \simeq \Delta \mathrm{Vol}
\label{eq:volume decomposition}
\end{equation}
where $s_j$ denote the surface areas of the faces of $M_0$.

\subsubsection*{Computation of local displacements} 
We estimate local displacement $d_{j}$ of a triangular face $j$ of mesh $M_0$ with respect to mesh $M_1$ by averaging the projection of the displacement vectors  $\vec{q}_{i}$ onto the face normal $\vec{n}_{j}$ . The average is taken over the vertices $i$ that form face $j$. Displacement vectors $\vec{q}_i$ are defined as $\vec{q}_i = \mu(\vec{v}_i) - \vec{v}_i$, where $\mu$ is the optimal matching between $M_0$ and $M_1$. The optimal matching $\mu$ is found by minimizing the sum of distances between pairs of matched vertices, $\mu = \mathrm{argmin}_{m}\sum_i ||m(v_i) - v_i|| $, over all possible one-to-one matchings $m$ between the vertices of $M_0$ and $M_1$

In summary, the local displacements of a mesh $M_0$ with respect to a mesh $M_1$ are computed as follows:

\begin{enumerate}
\item Find the matching $\mu$ that minimizes the total distance:
\begin{equation}
\mu = \mathrm{argmin}_{m}\sum_i ||m(v_i) - v_i||	
\label{eq: matching}
\end{equation}
where $m:\mathcal{V}_0 \to \mathcal{V}_1$ is a one-to-one correspondence between the vertices of $M_0$ and those of $M_1$. If $|\mathcal{V}_0| \neq |\mathcal{V}_1|$, we resort to subsampling the mesh with the largest number of vertices.  We solve Eq.~\ref{eq: matching} using the Hungarian algorithm \cite{Kuhn2005} as implemented by the \emph{scipy.optimize.linear\_sum\_assignment} function from the SciPy library~\cite{Jones2001}

\item Compute the displacement vectors $\vec{q}_i$ of each vertex:
$$
\vec{q}_i = \mu(v_i) - v_i
$$
\item Compute the normal vectors to all faces $\vec{n}_{j}$
\item Compute the projections of $q_{i}$ on $n_{j}$
\item The local displacement $d_{j}$ is the average of $q_{i}\cdot n_{j}$ over the vertices $i$ that form face $j$:

\begin{equation}
	d_j =  \vec{n}_j \cdot \left(\frac{1}{3} \sum_{i=0}^{2} \vec{q}_i \right)
\end{equation}

\end{enumerate}

\subsubsection*{Local displacement fluctuations}
The quantity $\sigma(d)$ reported in Figure \ref{fig:4} is the standard deviation of the local displacements $d_j$ of each cell. 
\begin{equation}
\sigma(d) = \sqrt{ \frac{1}{N}\sum_j d_j^2 -\left(\frac{1}{N} \sum_j d_j\right)^2} 
\end{equation}
large values of $\sigma(d)$ are associated with heterogenous morphological, while low values of $\sigma(d)$ are obtained when the changes are homogeneous over the nucleus surface. In particular, volume changes due to a uniform shrinking/expansion would entail a value of $\sigma(d)\ll 1$.

\subsection*{Statistical significance tests}
The $p$-values reported in Figure 1 and panels b) and e) of Figure SX are obtained via a Kolmogorov-Smirnov test comparing the distribution of the relevant magnitud among the two groups (Doxy vs No Doxy or Healthy vs HGPS). The computation is implemented via ``ks\_2samp'' function from SciPy library \cite{SciPy}. The p-values reported in panels c) and f) of Figure SX are instead computed using a T-test for independent samples with unequal sample sizes and unequal variances, implemented throught the ``ttest\_ind''function from SciPy library \cite{SciPy}. We use the Kolmogorov-Smirnov test to assess the statistical significance of differences in local displacements when comparing treated versus untreated cells. In Figure \ref{fig:4} $p$-values below 0.05 are marked with $^*$, and $p$-values below 0.01 are marked with $^{**}$. For each time-step, we correct $p$-values for multiple testing using the Holm-Sidak method as implemented in the ``statsmodel.stats.multites'' python library.
We also used t-test when indicated. 

	                                                                                                                                                                                                               \subsection*{Computational modeling}
We consider a discretized model for the cell nucleus composed by a flexible coarse-grained shell representing nuclear envelope and lamina, coupled to a set of coarse-grained polymers representing chromatin and also to a set of randomly distributed points representing cytoskeletal elements. At this level of coarse-graining, the intention is to capture the mechanical properties of the essential nuclear components without resorting to detailed modeling of their structure. 
A sketch of the model is reported in Fig. \ref{fig:sketch} and the list of parameters
is summarized in Table \ref{table:parameters}.

\paragraph{Chromatin:}  Chromatin is modeled in analogy with previous coarse-grained simulation studies \cite{Stephens2017,Stephens2018,Naumova2013,Ozer2015}. We consider a set of 46 polymers, with 128 monomers each. Monomers are coupled by harmonic springs with spring constant $10^{-3}$~N/m\cite{Stephens2017,Stephens2018}, and a harmonic angular coupling with equilibrium angle 130 degrees and spring constant $2\times 10^{-16}$ J$\cdot$rad$^{-2}$. To avoid that chromatin 
overlaps, monomers also interact with a truncated Lennard-Jones interaction with $\epsilon=10^{-3}$~pJ, $\sigma=0.12$~$\mu$m and a cutoff of $0.2$~$\mu$m, which provide an effective short-range repulsion. Chromatin polymers are initialized as non-overlapping random walks, each centered on a random site within the spherical nucleus. This initial condition ensures the presence of chromatin segregation into separated territories, as suggested by experiments \cite{meaburn2007chromosome}. The initial chromatin configuration is first relaxed using an NVE integrator with particle step size limited to 1~nm, per time step of $10^{-9}$~s, for $5\times10^4$ time steps. The system is then brought to thermal equilibrium using an NVE integrator with Langevin thermostat with target temperature $300$~K and damping $\gamma=10^6$~s$^{-1}$, with a time step of $10^{-7}$~$\mu$s, for $5\times10^7$ steps, before the nuclear envelope and cytoskeleton are added to the simulation.

\paragraph{Nuclear envelope:}
The nuclear envelope is modeled as a triangulated sphere, using 10242 nodes in the triangulation, and each node is treated as a particle in a molecular dynamics simulation.
The energy cost of stretching the shell is implemented by coupling each node of the triangulation by linear springs to its neighbors with spring constant $5\times10^{-3}$~N m$^{-1}$~\cite{Vaziri2007} for all bonds.  Bending the shell also has an energy cost, which is implemented by coupling each triangular plaquette of the sphere to its neighbors using a harmonic coupling in the angle between two plaquettes, with equilibrium angle 180 degrees. We use a spring constant for this coupling of $10^{-15}$ J rad$^{-2}$  inside the domains and $10^{-17}$ J rad$^{-2}$ on the domain walls. This is larger than the literature value of  $10^{-19}$ J rad$^{-2}$~\cite{Vaziri2007}, but we find this is necessary to prevent crumpling of the shell. The nuclear envelope is initialized as a sphere of radius $6$ $\mu$m. To prevent chromatin polymers from passing through the shell, the chromatin and shell particles interact with a truncated Lennard-Jones interaction with $\epsilon=10^{-3}$~pJ, $\sigma=0.6$~$\mu$m and a cutoff radius of 
$r_c=0.6$~$\mu$m. When the nuclear envelope is created, monomers in the relaxed chromatin configuration that are within distance $0.35$ $\mu$m of the nuclear envelope are able to couple to the nuclear envelope with probability $p_0$ that we can fine tune
to obtain a given lamin/chromatin link density $p$, defined as the average number of links
per node of the triangulated shell.

\paragraph{Lamin domains:}
A key feature of our model is the presence of spatial domains in the lamina, which are bounded by ``domain walls''  that are easier to bend than the domains. The existence of spatial domains was demonstrated experimentally in \cite{dahl2006} and we hypothesize that the domain walls are more bendable than the domains themselves.  The nodes are assigned to domains by generating a domain pattern on the surface of the sphere by placing 10 domain centers uniformly at random and then generating the domain walls using Voronoi tessellation.  When the nuclear envelope is created, monomers in the relaxed chromatin configuration that are within distance $0.35$ $\mu$m of the nuclear envelope are able to couple to the nuclear envelope only if the envelope particle is on a domain wall. In all cases, coupling is implemented via harmonic springs with stiffness 
$k_{tether}=10^{-2}$N/m. 

\paragraph{Coupling to the cytoskeleton:}
The nuclear envelope is also coupled to a set of 100 randomly placed points representing elements of the cytoskeleton. The coupling is implemented via linear springs that connect a randomly-selected points on the lamina, with spring constant $k_{cyto}$, that we vary in the range $10^{-3}-10^{-2}$~N m$^{-1}$. To simulate cytoskeletal activity, each cytoskeletal element 
follows an oscillatory motion $\vec{R}_i=\vec{R}_i^0-\vec{A}_i \cos(\omega_i t)$ with 
frequencies randomly distributed in the range $[0.05,0.2]\mu$s$^{-1}$ and amplitudes 
also randomly distributed in the range $[-1,1]\mu$m.

\paragraph{Simulation details:}
The system is first relaxed using and NVE integrator with particle step size limited to 1~nm, per time step of $10^{-9}$~s, for $10^6$ time steps. The system is then brought to thermal equilibrium using an NVE integrator with Langevin thermostat with target temperature $300$~K and damping constant $\gamma=10^{4}$~$\mu$s$^{-1}$, with a time step of $10^{-6}$~$\mu$s, for $6\times10^7$ steps. All simulations are implemented in LAMMPS \cite{plimpton1995fast}.

\section*{Results and discussion}

\subsection*{Progerin induction in HeLa cells modifies nuclear morphology and chromatin organization}
The Tet-On system is based on reverse tetracyclin activation. Here we used the Tet-On 3G system which is 100 fold more sensitive and 7 fold more active than the original Tet-On (see  Material and Methods section for technical information). In our cellular model, the level of expression of progerin is induced  by doxycycline (Doxy). As shown in Fig. \ref{Fig:S1}, 10ng/ml Doxy is the minimal concentration that is able to induce progerin expression. To check the specificity of the induction of progerin after the treatment with Doxy, we used two internal controls according to the manufacturer's instructions: we compared the cells treated with  Doxy with respect to cells not exposed to Doxy (No Doxy). The second control was  to investigate the level of expression of progerin in cells transfected  with an empty vector, untreated or after treatment with Doxy 
(Fig. \ref{Fig:S1}C). In all these control experiments, we did not observe any 
progerin induction. For completeness in Fig. \ref{Fig:S2_new}, we show the mCherry expression
and panlamin for the same conditions as in Fig. \ref{Fig:S1} and in Fig. \ref{fig:emerald}
an example of HeLa cells expressing mEmerald and WT lamin A or pEGFP-$\Delta$50 lamin A  48h after transfection. We also quantified by western blot the level of expression of lamin A and B under the induction of progerin, without finding any significant changes due to the presence of progerin  (Fig. \ref{fig:S2}).

It was reported in the literature that in cells from HGPS patients, the ratio between the amount
of progerin and lamin A proteins is between 0.5 and 2
\cite{Reunert2012,Vidak2015,dahl2006}. This variability is possibly due to the specificity of each patient. Here, we quantified the ratio between progerin and lamin A proteins in HeLa Tet-On progerin expressing cell and found a value around 1.5-2, as shown in Fig. \ref{fig:S3}.  

\begin{figure*}[h]
\begin{center}
\includegraphics[width=15cm]{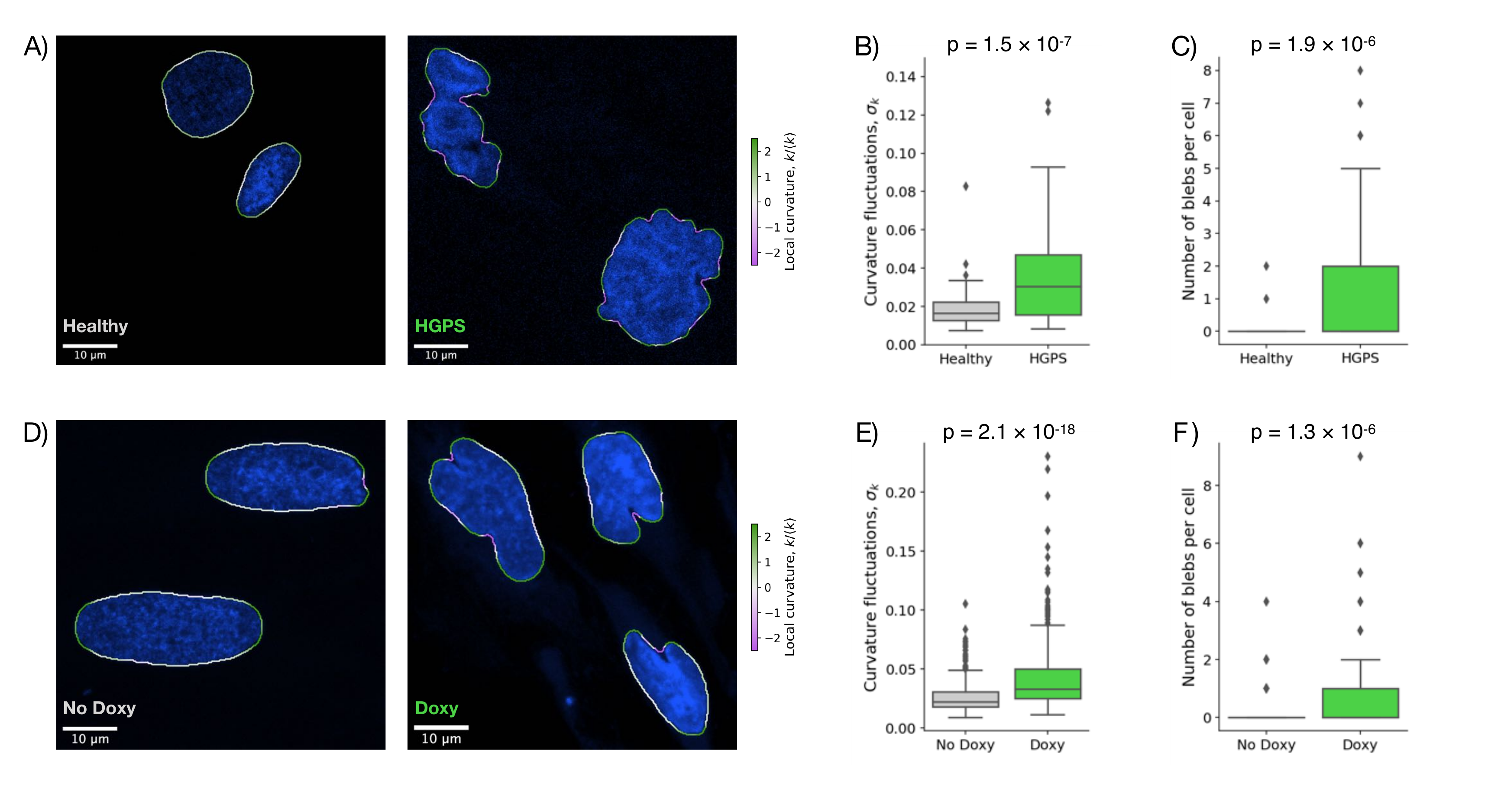}
\end{center}
\caption{{\bf Progerin expression affects nuclear morphology}.  Typical nuclei obtained from A) subconfluent HGPS cells (HGPS), healthy mother of HGPS patient  (Healthy),  or D) HeLa Tet-On progerin expressing cells without Doxy tratment (No Doxy) or after induction with 10ng/ml with Doxycyclin (Doxy). Subconfluent cells were fixed with ice-cold methanol, then incubated anti-PanLamin (1:50, ab20740, Abcam) at 4$^\circ$C overnight and with AlexaFluo488 (1:250, ab15113, Abcam) for 1h, at RT. Nuclei were stained with DAPI. Images were acquired by Leica SP2 laser scanning confocal microscope. Quantification of morphological alterations of B-C) HGPS cells, mother of HGPS cells, E-F) HeLa Tet ON progerin expressing cells (Doxy) and HeLa Tet-On cells without Doxy treatment (no Doxy)  by computing the number of blebs (B and E) and the curvature fluctuations (C and F) as described in the material and method section. Statistics was performed over 297 No Doxy nuclei, 330 Doxy nuclei, 87 healthy nuclei, and 87 HGPS nuclei.The $p$-values reported are obtained via a Kolmogorov-Smirnov test comparing the distribution of the relevant magnitude among the two groups (Healthy vs HGPS or Doxy vs No Doxy). The computation is implemented via ``ks\_2samp'' function from SciPy library \cite{SciPy}. Data have been
collected over at least three independent experiments. 
\label{fig:1}}
\end{figure*}

To confirm that our cellular model faithfully represents the morphology of HGPS nuclei, 
we compared the number of blebs and the curvature fluctuations observed in HeLa Tet-On progerin expressing cells (Doxy) with those observed in control cells  (no Doxy) and performed a similar
analysis for fibroblasts obtained from a HGPS patient (HGPS) and his healthy mother (Healthy) (Fig.~\ref{fig:1}).  Using the same samples, we also measured a blebbing index, defined as the average curvature restricted to the region of negative curvature (see Materials and Methods section for more details) and the number of cells with blebs  (see Fig.~\ref{fig:S4}). All these different quantitative measurements confirmed that nuclei 
in both HeLa Tet-On progerin expressing cells and HGPS cells display 
highly significant alterations with respect to their controls, confirming the capability of our model to faithfully recapitulate the main features of the nuclei from HGPS patients. 
As a further control, we report in Fig. \ref{fig:empty} morphological measurements
in cells transfected with an empty vector. In this case, treatment with Doxy does not
induce any morphological alteration. 

Finally, it is known that HGPS patients show important differences in higher order chromatin organization such as specific changes in chromatin-modifying enzymes (i.e. heterochromatin protein 1, HP1) \cite{scaffidi2008,gabriel2015}. Consistently,   we found an increased expression of HP1 in HeLa Tet-On progerin expressing cells (Fig. \ref{fig:S5}).
 
\subsection*{Progerin induction affects the coupling between the cytoskeleton and the nuclear shell}

\begin{figure*}
\begin{center}
\includegraphics[width=15cm]{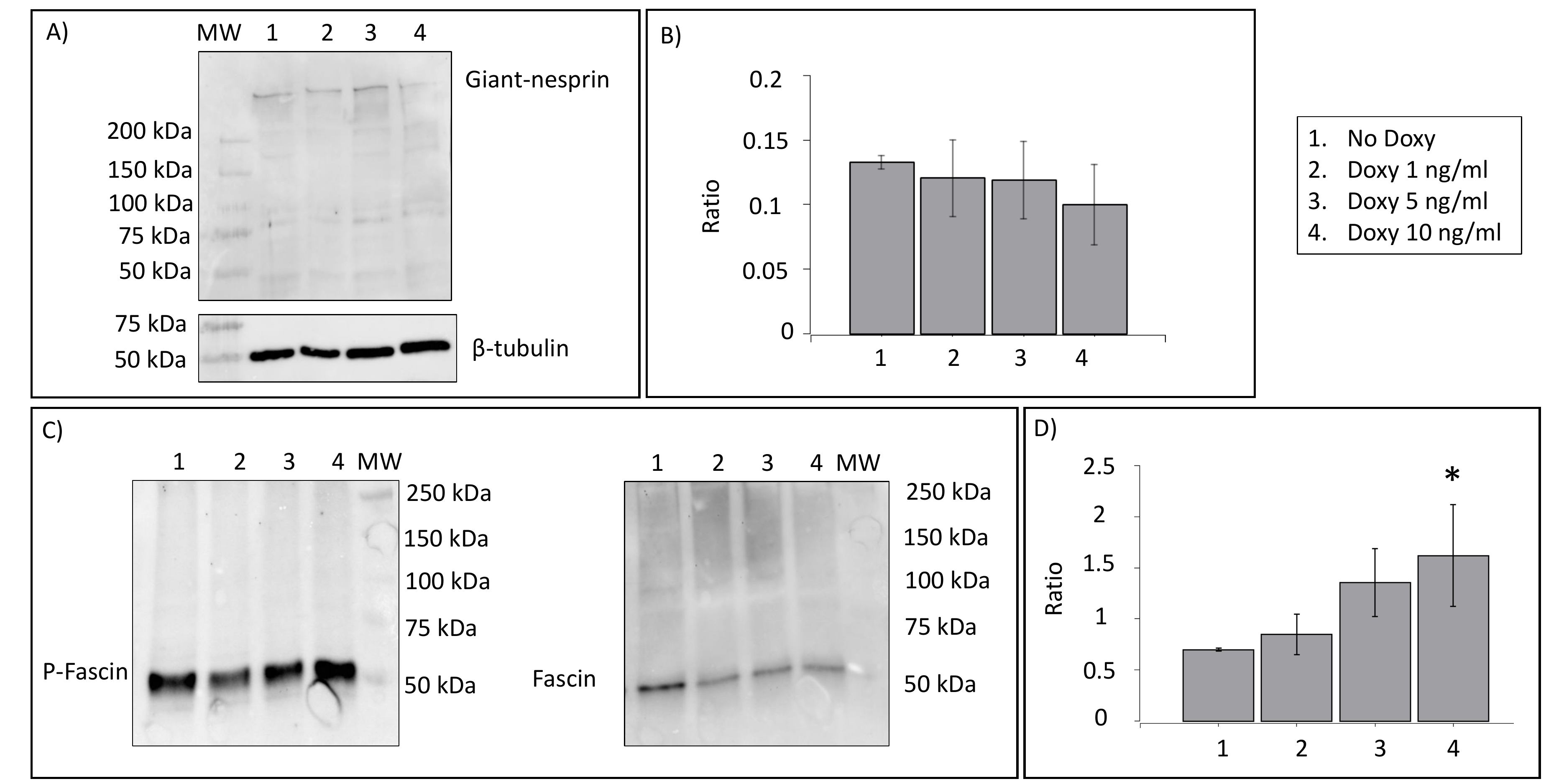}
\end{center}
\caption{{\bf Progerin induction leads to fascin phosphorilation in S39 without affecting nesprin-2 expression}. A) Typical experiment of western blot of giant nesprin in HeLa Tet-On 3G progerin expressing cells treated with increasing concnetration of Doxycyclin (Doxy) or without (No Doxy). 20$\mu$g total protein were loaded on 10\% polyacrylamide gel, transferred on PVDF and incubated with nesprin-2 (1:1000, MABC86, Merck Millipore) overnight at 4$^\circ$C. anti-$\beta$-tubulin antibody (1:5000, T8328, Sigma ) for 1 h at RT was used as housekeeping. B) Densitometric analysis of two independent western blot experiments of giant nesprin western blot as shown in Panel A.  Densitometric analysis was carried out with ImageJ software. The Y axis shows the ratio between the mean of the densitometric value of giant nesprin with respect to the housekeeping $\beta$-tubulin. Statistical significance was established by the t-test. C) Typical western blot of S39-fascin  in  fascin-immunoprecipitated samples obtained from HeLa Tet-On progerin expressing cells treated or untreated with Doxy. Subconfluent cells were processed for immunoprecipitation as described in the Materials and Method section.  Briefly,  500 $\mu$g of total proteins was incubated with anti-fascin antibody (1:100,  ab126772, AbCam)  overnight at 4$^\circ$C under stirring.  50\% beads slurry of Protein A-Agarose (P9269, Sigma Aldrich) was added to the lysate and reincubated with gentle rocking for 2 hours at  4$^\circ$C. After three washes with 500$\mu$l of lysis buffer the sample was resuspended in 30$\mu$l 2X Laemmli sample buffer, heated to 90$^\circ$C for 5 minutes, 15 $\mu$l of immunoprecipitated sample was loaded on 10\% SDS-PAGE for Western blot. D) Densitometric analysis of two independent experiments as shown in panel C. The Y axis shows the ratio between the mean of the densitometric value of S39-fascin and fascin. 
\label{fig:2}}
\end{figure*}

Since the interaction between fascin and giant nesprin or nesprin-2 requires the phosphorylation of fascin at serine 39 (S39),  we investigated the possible impairment of fascin/nesprin-2 complex in HeLa Tet-On progerin expressing cells. As shown in Fig. \ref{fig:S6}, nesprin-2 associates with the outer nuclear membrane exposing their N-term domain
towards the cytoplasm where it binds cytoskeletal actin through fascin and their C-term domain towards the perinuclear space where it binds to SUN1/2 and emerin. According to Ref. \cite{Jayo2016}, fascin can interact properly with nesprin-2 only when it is phosphorylated (S39).
First, we checked the level of expression of giant nesprin in HeLa Tet-On progerin expressing cells with respect to cells not induced by Doxy. As shown in  Fig.~\ref{fig:2}, the induction of progerin did not change the level of expression of nesprin-2. On the other hand, we observed an increase in the expression level of S39-fascin as progerin level of expression increases (Fig.~\ref{fig:2}).  Hence, our observation suggests that the expression of progerin enhances the coupling between the cytoskeleton and the nuclear shell (Fig.~\ref{fig:2} and Fig. \ref{fig:S6}). 

\subsection*{Progerin induction affects chromatin-lamin tethers}
\begin{figure*}
\begin{center}
\includegraphics[width=15cm]{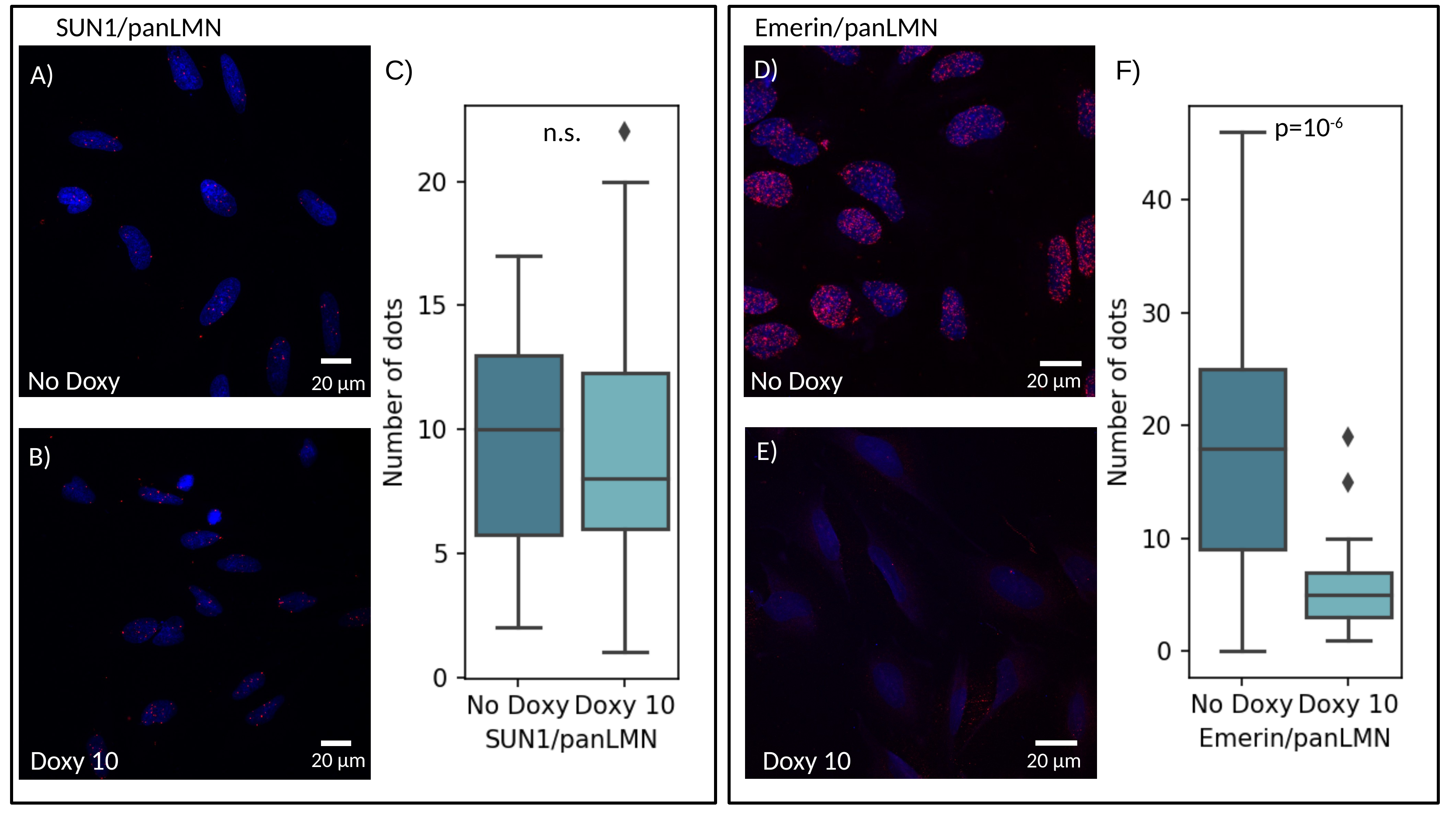}
\end{center}
\caption{{\bf Effect of progerin induction on nuclear tethering factors by proximity ligation assay.} The tethering is quantified by  the proximity ligation assay measuring the interaction between all lamins detected using PanLamin antibody and (A-B) SUN1 or (D-E) emerin. Briefly, subconfluent cells were fixed on slides with ice cold 100\% methanol for 5 min. Slides were then incubated  in an humidity chamber overnight at 4$^\circ$C with  PanLamin (1: 50, ab20740, Abcam) antibody and with SUN1 (1:200, ab103021, Abcam) or emerin (1:200, ab40688 Abcam). After washing, the samples were incubated in a pre-heated humidity chamber for 1 hour at 37$^\circ$C with anti-rabbit PLUS and anti-mouse MINUS PLA probes diluted 1:5.  Ligation and amplification steps were performed according to the manufacturer's instructions.  Slides were mounted with Duolink In Situ Medium. The number of aggregates linking C) SUN1 or F) emerin and lamins are quantified as described in the Materials and Methods section. The analysis has been carried out on 72 nuclei
for SUN1 (36 without Doxy and 36 with Doxy) and 76 nuclei for emerin (53 without Doxy and 23 with Doxy). Statistical significance is established by the Kolmogorov-Smirnov method. Data have been
collected over at least three independent experiments. 
 \label{fig:3}}
\end{figure*}

To study the role of progerin induction in the tethering between lamins and chromatin, we investigated the presence and the interaction between SUN1 or emerin with all lamins using the PanLamin antibody in HeLa Tet-On progerin expressing cells. 
Fig. \ref{fig:S7}, shows that neither SUN1 nor emerin level of expression changes significantly in dependence on the expression of progerin. We then investigated the dependence on progerin expression of the interaction between lamins and SUN1 or emerin using the proximity ligation assay. In Fig. \ref{fig:3}, we show a typical experiment and its quantification by counting the number of the spots, each representing an interaction between SUN1 or emerin and lamins, as detected with the PanLamin antibody. Fig. \ref{fig:3} shows a significant reduction in the coupling between emerin and lamins upon progerin induction. In contrast, the coupling between SUN1 and lamins is unaffected by progerin induction Fig. \ref{fig:3}.

\subsection*{Progerin induction slows down collective cell migration}
To assess the effect of progerin induction on functional properties of the cell, we performed wound healing assays. As shown in Fig. \ref{fig:S8}A, the velocity distributions appears altered by progerin induction leading to a slower collective migration. While the effect is small it is statistically significant according to the Kolmogorov-Smirnov test. Accordingly, the front position of progerin induced cells advances more slowly than in control cells (see  Fig. \ref{fig:S8}B). 

\subsection*{Nuclear surface fluctuations driven by cytoskeletal activity are affected by progerin}

\begin{figure*}
\begin{center}
\includegraphics[width=10cm]{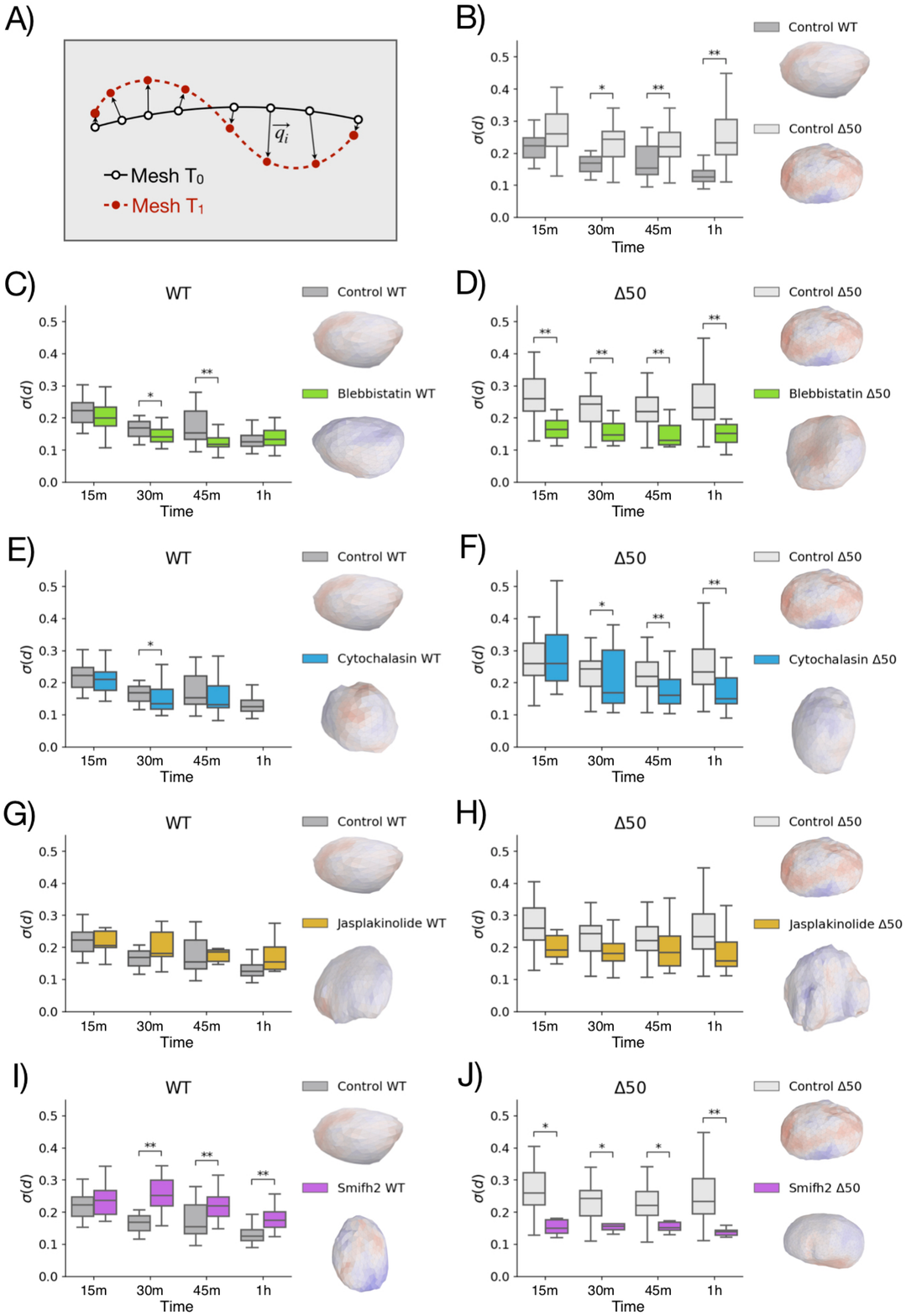}
\end{center}
\caption{{\bf Progerin induction leads to larger shape fluctuations due to cytoskeletal activity.} A) Illustration of local displacements method: given meshes for two timepoints $T_0, T_1$, the algorithm finds the vectors $v_i$ used to compute $\sigma(d)$, see Methods for details. 
B) Boxplot of $\sigma(d)$ over time for untreated cells, comparing cells overexpressing WT lamin A (dark grey) or $\Delta$50 lamin A (light grey). Panels C-D show
boxplots of $\sigma(d)$ over time comparing  control cells (gray) with blebbistatin-treated cells (green), for C) WT lamin A and D) progerin cases. Additonal panels show  cells treated with E-F) cytochalasin (blue boxplots), G-H) jasplakinolide (orange boxplots) and I-J) SMIFH2 (violet boxplots). For each time point, local displacements are computed with respect to the previous time point. Statistical significance is measured with Kolmogorov-Smirnov tests, with ** (*) marking $p<0.01 (0.05)$. $p$-values are corrected for multiple testing, see Methods for details.
Data have been
collected over at least three independent experiments. 
\label{fig:4}
  }
\end{figure*}

Forces generated by the cytoskeleton can affect chromatin through the links with the nuclear envelope and the lamina, as illustrated in Fig. \ref{fig:S6}. To quantify the effect of cytoskeletal activity on the nuclear envelope when progerin is expressed, we studied time-dependent shape fluctuations in HeLa cells transiently transfected with  GFP---$\Delta$50 lamin A and mEMERALD-lamin A WT. To this end, we exposed the cells to agents that are able to stabilize actin (jasplakinolide), depolymerize actin (cytochalasin D), affects myosin activity (blebbistatin) or inhibit the nucleation of actin polymerization (SMIFH2) and then we wash out the drug and starts to acquire images at confocal microscope \cite{Makhija2016}.   We thus reconstructed the 3D mesh of nuclear surfaces according to Materials and Methods section and  from each time-series of nuclear meshes, we computed the local displacement fields $d_i$, indicating how much each node of the mesh has been displaced from the beginning (See Fig. \ref{fig:4}A). Finally, to quantify the fluctuations of the mesh in each instance, we computed the standard deviation of $d_i$ as a function of time. As shown in Fig. \ref{fig:4}B, progerin overexpression leads to significantly larger time-dependent fluctuations in the nuclear envelope. Notice that larger fluctuations is not trivially related to the presence of blebs. Fluctuations here are defined relative to a reference frame, so that a blebbed surface not changing in time would yield no fluctuations. Moreover, the treatments with blebbistatin 
(Fig. \ref{fig:4}CD) and SMIFH2 (Fig. \ref{fig:4}IJ)  and with less extend to jasplakinolide
(Fig. \ref{fig:4}GH), showed significant differences in the presence of progerin. The general conclusion is that hindering cytoskeletal activity affects progerin expressing cells reducing their nuclear surface fluctuations.

\subsection*{Progerin affects the interaction between lamins and polycomb proteins}
\begin{figure*}
\begin{center}
\includegraphics[width=12cm]{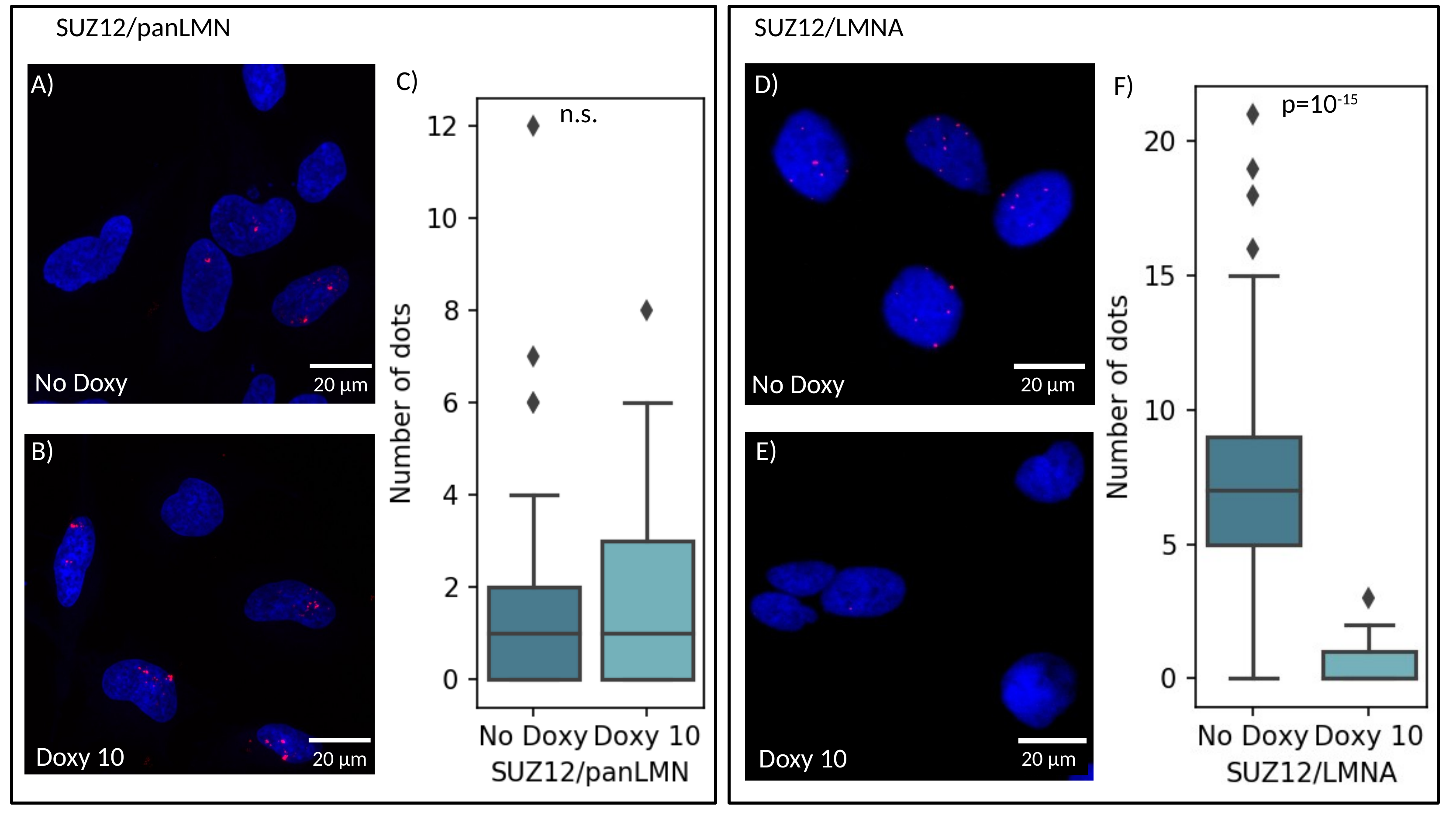}
\end{center}
\caption{\label{fig:5} {\bf Effect of progerin induction on the interaction of polycomb protein SUZ12 and lamins by proximity ligation assay.} 
The tethering is quantified by the proximity ligation assay measuring the interaction between A-B) all lamins (panLMN) or D-E), lamin A  and PcG SUZ12. Briefly, subconfluent cells were fixed on slides with ice cold 100\% methanol for 5 min. Slides were then incubated  in a humidity chamber overnight at 4$^\circ$C with  PanLamin (1: 50, ab207404, Abcam) or anti-lamin (1:100 ab8980, Abcam)  antibody with  SUZ12 (1:800,mAb 3737 Cell Signaling). After washing, samples were incubated in a pre-heated humidity chamber for 1 hour at 37$^\circ$C with anti-rabbit PLUS and anti-mouse MINUS PLA probes diluted 1:5.   Ligation and amplification steps were performed according to manufacturer's instructions.  Slides were mounted with Duolink In Situ Medium. Panels A), B), D) and E) show typical experiments for each experimental conditions. The number of aggregates linking SUZ12 and lamins (C for PanLamin and F for lamin A) are quantified as described in the Materials and Methods section. The analysis was carried out on 79  nuclei for PanLamin (33 without Doxy and 46 with Doxy) and 81 for lamin A (41 without Doxy and 40 with Doxy). Statistical significance is established by the Kolmogorov-Smirnov method. Data have been
collected over at least three independent experiments. }
\end{figure*}

Polycomb Group (PcG) proteins play an important role in chromatin remodeling during development. High-throughput data combined with microscopy analysis revealed a specific organization of their targets in chromatin loops \cite{Cmarko2003,Bantignies2011} that is altered in lamin A null cells
\cite{Bantignies2011}. To investigate the interactions between PcG proteins and lamins, we carried out proximity ligation assay. According to this technique, a fluorescent signal is detectable when two proteins are in close proximity (less than 40nm). First, we checked which PcG were expressed in Tet-On Hela cells without the Doxy treatment and after the induction of progerin.  As shown in Fig.~\ref{fig:S9}, we detected the presence of EZH2, SUZ12, Ring1A and BMI1. In contrast, Ring1B was not expressed in a detectable way. We carried out Proximity ligation assay for the two PcGs  belonging to the two PcG machineries: BMI1 (PcG1) and SUZ12 (PcG2) which are involved in chromatin remodeling through two distinct mechanisms \cite{Chittock2017} with all the lamins using the PanLamin antibody or with lamin A. As shown in Fig.~\ref{fig:5},  the interaction between SUZ12 and lamin A is significantly changed by the induction of progerin, while no effect is revealed for BMI1 (Fig.~\ref{fig:S10}).

\subsection*{Computational model indicates the relevance of tethering on lamin domains walls for nuclear bleb formation} 

\begin{figure*}
\begin{center}
\includegraphics[width=12cm]{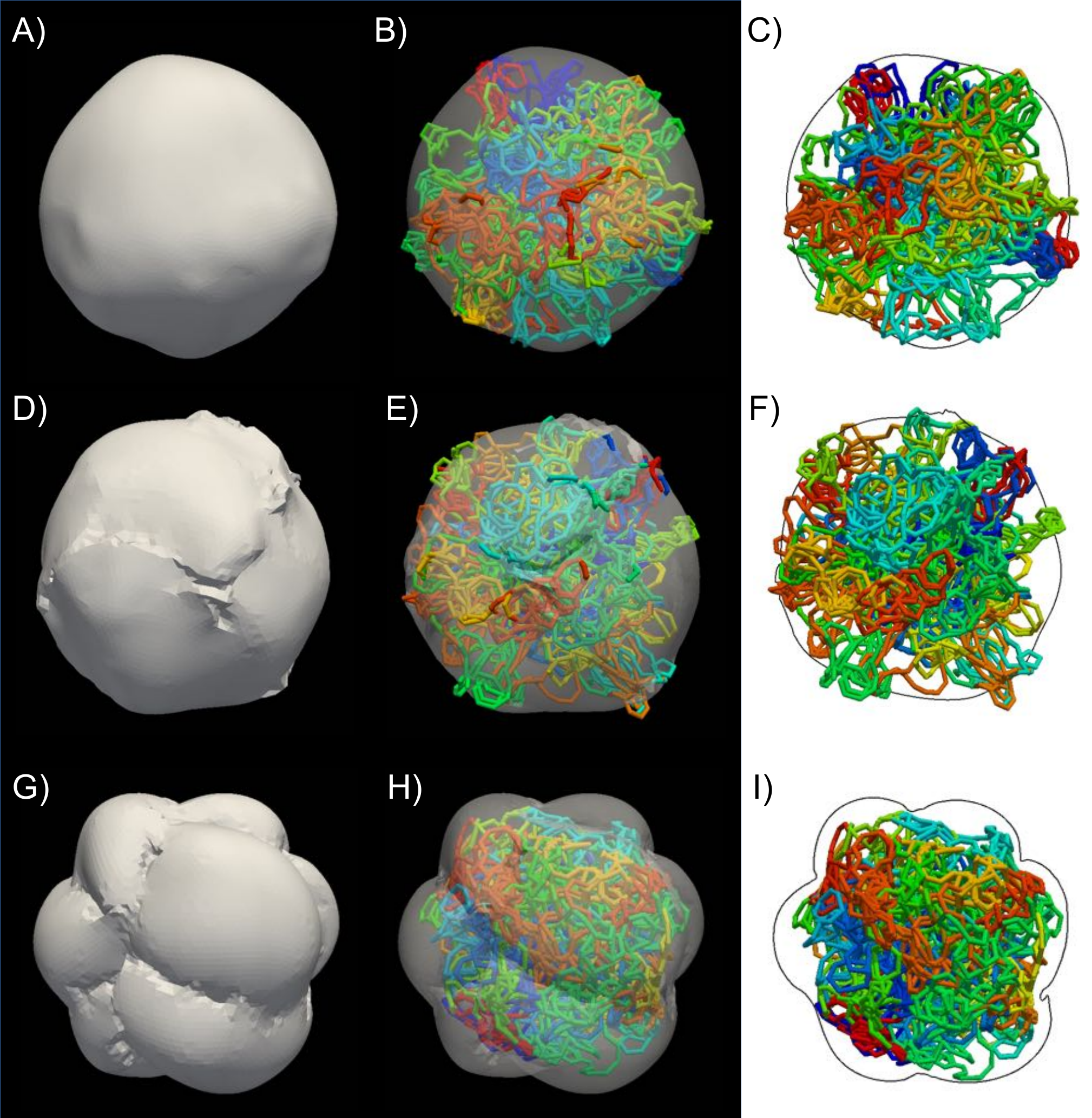}
\end{center}
\caption{{\bf Formation of nuclear blebs is induced by a strong chromatin tethers localized on lamin domain boundaries}. A-B-C) Simulations without lamin domains and uniform distribution
of lamin-chromatin tether.  D-E-F) Simulations with lamin domains and uniform distribution
of lamin-chromatin tether. G-H-I) Simulations with lamin domains and lamin-chromatin
tethers localized along the domain boundaries. Only in the last case, we observe the
formation of blebs. In all cases, the link density is $p=0.4$.
\label{fig:6} }
\end{figure*}

To further investigate the role of chromatin/lamin tethering for nuclear morphological alteration, we performed numerical simulations according to the protocol discussed in the modeling section.   We first considered the case in which chromatin tethers were distributed uniformly on the nuclear envelope. As shown in Fig. \ref{fig:6}A-C, while surface deformations 
are present, no distinct blebs are observed. We then varied two key parameters: the density of tethers $\rho$ and their stiffness $k_{tether}$. As illustrated in Fig. \ref{fig:random_noblebs}, increasing the tether stiffness and their density leads to nuclear alterations. Yet, these 
morphological alterations are in the form of creases and crumples, that look qualitatively very different from blebs observed experimentally. 

Previous experimental results indicate that in HGPS nuclei lamin is organized into domains 
characterized by distinct orientations as revealed by polarized light microscopy \cite{dahl2006}.
In WT nuclei, lamin filaments have no preferred orientation while in HGPS 
nuclei, lamin filaments are ordered within distinct domains. To take this observation
into account, we introduced lamin domains in the model by defining domain boundaries with low bending stiffness, so that the nuclear shell is folding more easily along those regions. Even in this case, blebs were not observed (Fig. \ref{fig:6}D-E) and we recorded only minor alterations with respect to a similar configuration without domain boundaries (Fig. \ref{fig:6}A-C). 

We then investigated the effect of lamin domains on the coupling between lamins and chromatin.
Modifications of the coupling is suggested by the experiments reported in Fig. \ref{fig:3}D-F
showing that a net reduction of the interactions between lamin and emerin, a chromatin 
tethering factor. To incorporate this observation into the model, we notice that
lamins are disordered in WT nuclei, while they are ordered within domains in HGPS nuclei. Hence,
we it is reasonable to assume that chromatin/lamin coupling occurs at the domain wall boundaries
where lamin organization is less ordered. Indeed, when chromatin tethers were attached exclusively at the lamin domain boundaries, we were able to recover nuclear shapes with blebs similar to those encountered in experiments (Fig. \ref{fig:6}F-H). 

To confirm that tethering to lamin domain is essential to account for blebs, we considered alternative non-uniform distribution of the chromatin-lamin tethers. For instance in Fig. \ref{fig:regions_lines}A, we report simulations results obtained placing all the tether in a single octant of the envelope. No blebs were found, but only localized depressions. When tether are instead placed along a single line, the sphere was found to crumple in a way that is very different from the experimental images (Fig. \ref{fig:regions_lines}B). The model also allowed us to describe the role of the elasticity of the nuclear envelope. In particular, we find that when the bending stiffness is reduced the smooth blebs observed in Fig. \ref{fig:6}G give way to the crumpled surface shown in Fig. \ref{fig:crumple}. Finally,  we studied the effect of the stiffness of the chromatin tether on the bleb morphology and found that reducing the stiffness blebs were less pronounced (Fig. \ref{fig:S14}). In sum, our computational results show that the local organization of the chromatin-lamin tethers is the key factor controlling bleb formation.
In particular, lamin domain wall boundaries are pulled inside the nucleus by chromatin tethers, creating the folds required for bleb formation. 

\subsection*{Computational model confirms the role of cytoskeleton in nuclear shape fluctuations}
As discussed in the previous section, the computational models is able to reproduce the 
morphological alterations of HGPS nuclei by considering the combined effects of lamin domains and chromatin-lamin tethers. We thus decided to use the model to better understand why 
nuclear shape fluctuations are enhanced when progerin is overexpressed (see Fig. \ref{fig:4}B). 
We reported in Fig. \ref{fig:2}D that progerin induction leads to increased fascin phosphorilation, leading to more stable couplings between cytoskeleton and lamins. 
We thus incorporated this information into the model by varying the strength of the
coupling between cytoskeleton and the nuclear shell. 

\begin{figure*}
\begin{center}
\includegraphics[width=12cm]{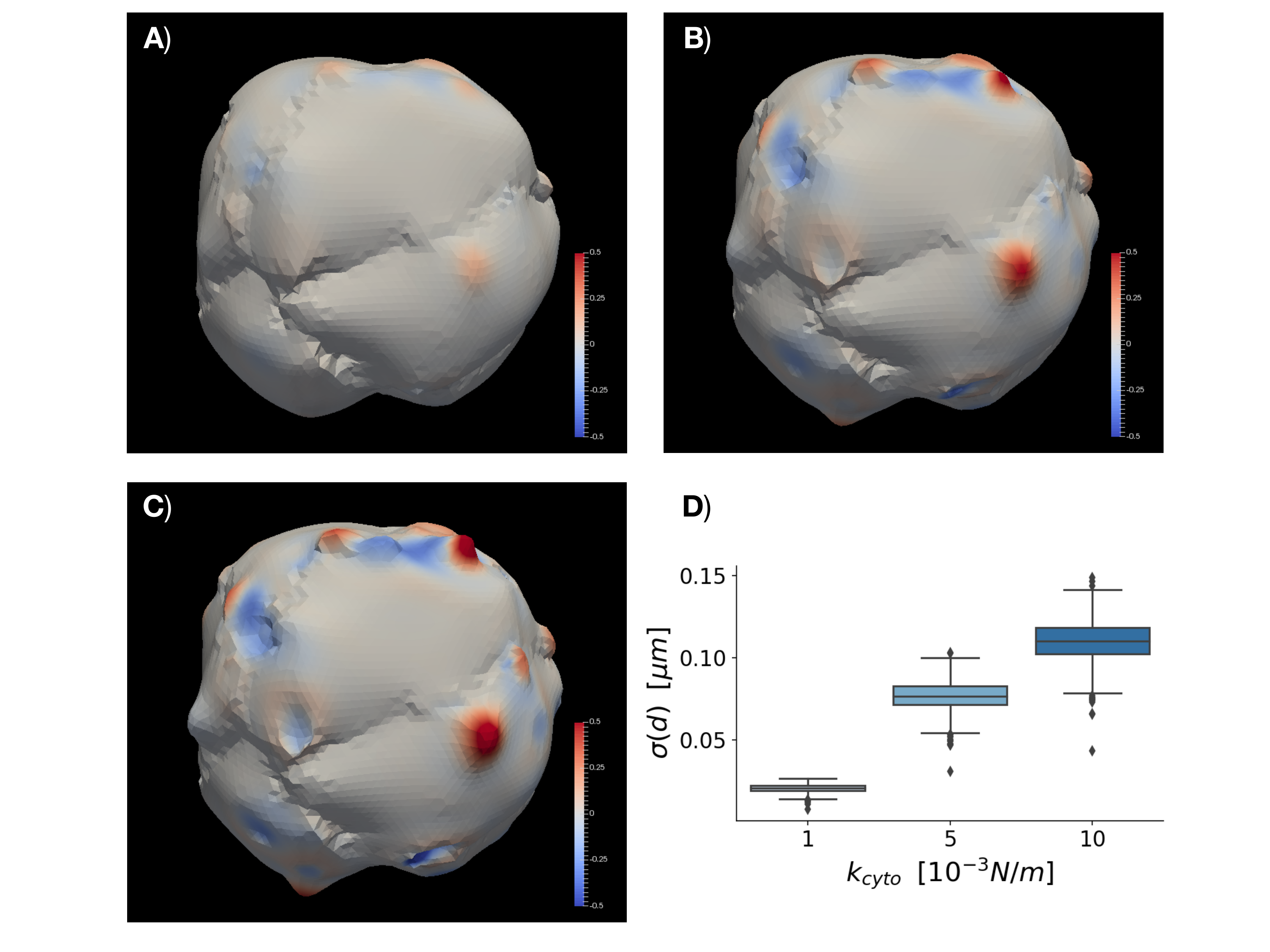}
\end{center}
\caption{{\bf Nuclear shape fluctuations depend on cytoskeletal tethering stiffness.} 
Simulations of shape fluctuations induced by cytoskeletal contraction for different
values of the stiffness of the tether $k_{cyto}$. A) $k_{cyto}= 10^{-3}$N/m, B) 
$k_{cyto}= 5 \cdot 10^{-3}$N/m C) $k_{cyto}= 10^{-2}$N/m. D) The standard deviation of
the radial displacements as a function of $k_{cyto}$. All the comparisons are 
statistically significant ($p < 10^{-10}$ according to the Kolmogorov-Smirnov test).
\label{fig:7} }
\end{figure*}

We simulated cytoskeletal contractions by a set of randomly placed 
oscillating points attached to the nuclear lamina by springs of stiffness $k_{cyto}$,
as described in the model section. Increased fascin phosphorilation was represented by
an increasing value of $k_{cyto}$. We then studied the fluctuations of the nuclear
shape induced by cytoskeletal activities (for an illustration of the fluctuations
see Fig. \ref{fig:7}A and Video S1). To quantify nuclear fluctuations, we computed 
the standard deviation $\sigma(d)$ of  the radial projection of the local displacement fields as a function of $k_{cyto}$, in close analogy with the 3D morphological analysis summarized
in Fig. \ref{fig:4}. The results reported in Fig. \ref{fig:7}B show that stronger
cytoskeletal coupling leads to enhanced fluctuations in good agreement with the results
reported in Fig. \ref{fig:4}B indicating that progerin overexpression is
associated with larger nuclear fluctuations.

\section*{Conclusion} 

In this paper, we introduced an inducible cellular model for progerin, the mutated form of lamin A responsible for HGPS. Through this cellular model, we analyzed how progerin affects nuclear tethering with chromatin and the cytoskeleton. Our results show that progerin expression leads to enhanced tethering between the cytoskeleton and the nuclear shell, through the phosphorilation of S39-fascin,  and at the same time to a reduced interaction between lamins and emerin, a known chromatin tether. These changes are reflected in alterations of the mechanical response of the nucleus. In particular, cytoskeletal-induced surface fluctuations of the nuclear shell are increased upon progerin expression. 

It has been shown that the coupling between the cytoskeleton and the nuclear shell can affect 
gene regulation \cite{Swift2014} by polycomb-mediated chromatin remodeling \cite{Le2016}. Hence, we studied the interaction between lamins and polycomb proteins upon progerin inductions. 
Our findings  show that progerin induction reduces the interactions between lamins and 
the polycomb protein SUZ12, suggesting progerin expression might impact on PRCs-mediated  gene expression regulation. 

Finally, we proposed a computational model for nuclear mechanics in which we simulate the conditions leading to nuclear morphological alterations, focusing in particular on the blebs observed in HGPS. Our results show that the presence of lamin domains is essential to recapitulate {\it in silico} the observed morphology. A crucial role in bleb formation is played by chromatin-lamin tethers: the presence of homogeneously distributed tethers suppresses blebs or leads to wrinkling of the surface. Blebs are only observed when tethering is concentrated at the boundaries of lamin domains. 

Our results are not obtained directly from HGPS cells but from an induced cellular model of progerin in HeLa cells that, however, recapitulates the main features of nuclear morphology and chromatin organization of HGPS cells, as shown for instance by the expression of HP1. Our cellular model overcomes many practical experimental limitations posed by the study of HGPS cells that are able to grow in vitro only for a few passages. 

All in all, our study highlights that progerin induction affects in crucial
ways the tethering between cytoskeleton, lamins and epigenetic regulation of chromatin
status. Simulations suggests that the strength and geometrical  organization of these tethers appear to control the morphological alteration observed in the nuclei of HGPS cells.  Our observations could be relevant for HGPS where therapeutic strategies should aim at restoring the cytoskeleton-lamin-chromatin coupling integrity.

\section*{Author Contributions}

MCL performed experiments. MRF, FFC, GC, OC, SZ analyzed data. ZB, SZ designed the
computational model. ZB, SB performed numerical simulations. CAMLP designed and 
coordinated the study. SZ and CAMLP wrote the paper.

\section*{Acknowledgments}
The research was supported by funding from the Center for Complexity and Biosystems
of the University of Milan. We thank C. Giampietro and F. Mutti for useful discussions
and assistance at the beginning of the project. 
S. Z. thanks the Alexander von Humboldt foundation for the Humboldt Research Award for 
support and Ludwig-Maximilian 
University and Friedrich-Alexander-Universit\"at Erlangen-N\"urnberg for hospitality.


\clearpage

\clearpage

\section*{Supplementary Material}

An online supplement to this article can be found by visiting BJ Online at \url{http://www.biophysj.org}.

\renewcommand{\thefigure}{S\arabic{figure}}
\setcounter{figure}{0}

\renewcommand{\thetable}{S\arabic{table}}
\setcounter{table}{0}

\begin{figure}
\centering
	\includegraphics[width=0.8\textwidth] {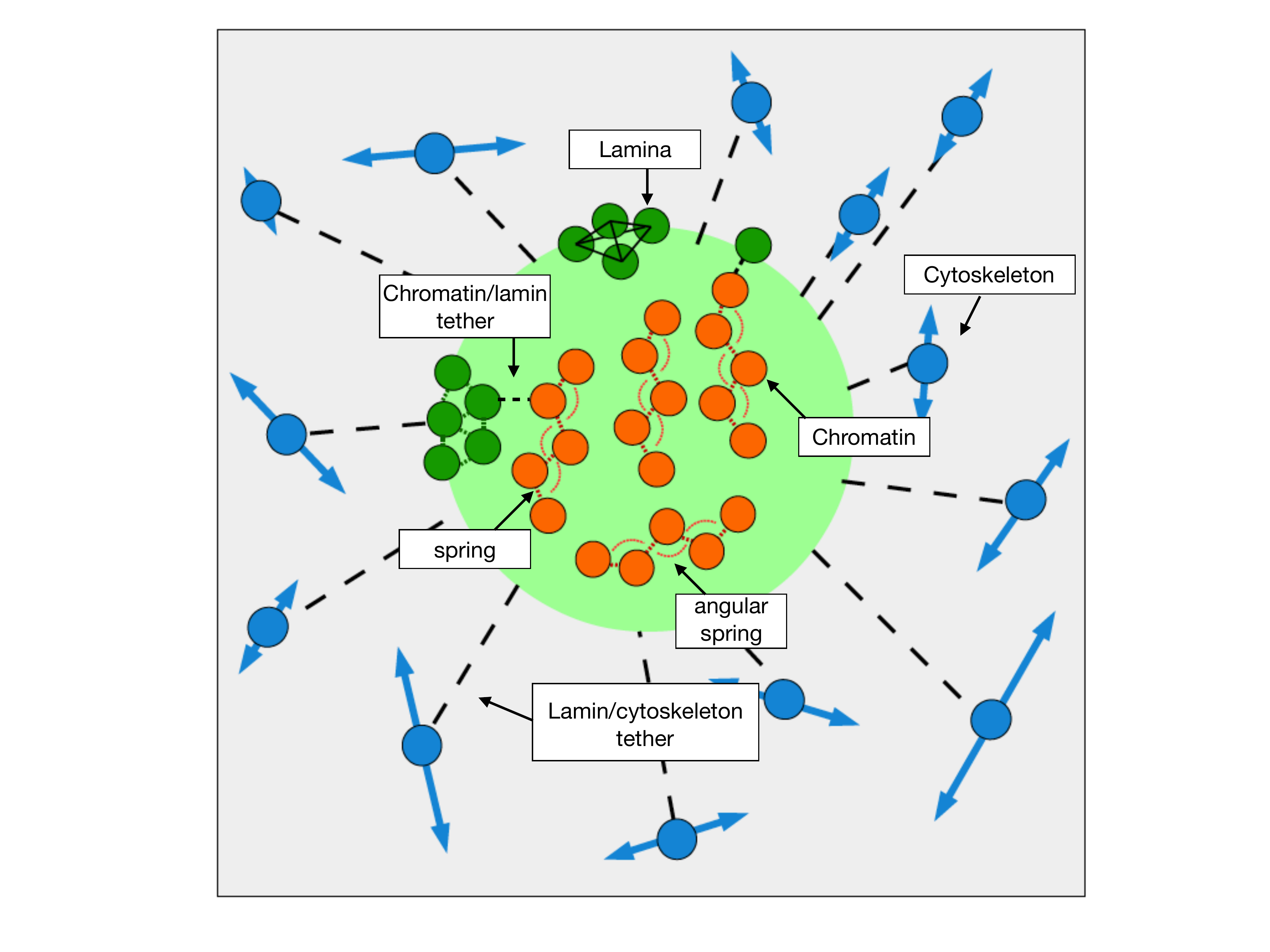}
	\caption{Sketch of the simulation setup: chromatin filaments inside the nuclear lamina surrounded by the cytoskeleton. The nuclear lamina (green region) is composed of green beads which interact with harmonic springs (green dashed lines) and impropers (upper 4 green beads in a tetrahedral shape). Monomers forming chromatin (orange beads) are  characterized by harmonic 
bonds and angular springs (dashed and dotted orange lines). Cytoskeletal activity is simulated by external beads (blue) which are bonded with nearest lamina beads, via harmonic bond. Harmonic coupling between lamin and chromatin is also present as written in the text with a specified probability (black dotted line between green and orange bead). Chromatin filaments interact with a truncated LJ, like the lamina particle and rest of chromatin filaments. Arrows with random amplitude and directions attached to blue beads represent the movement of cytoskeleton.}
	\label{fig:sketch}
\end{figure}

\begin{figure*}[h]
\begin{center}
\includegraphics[width=12cm]{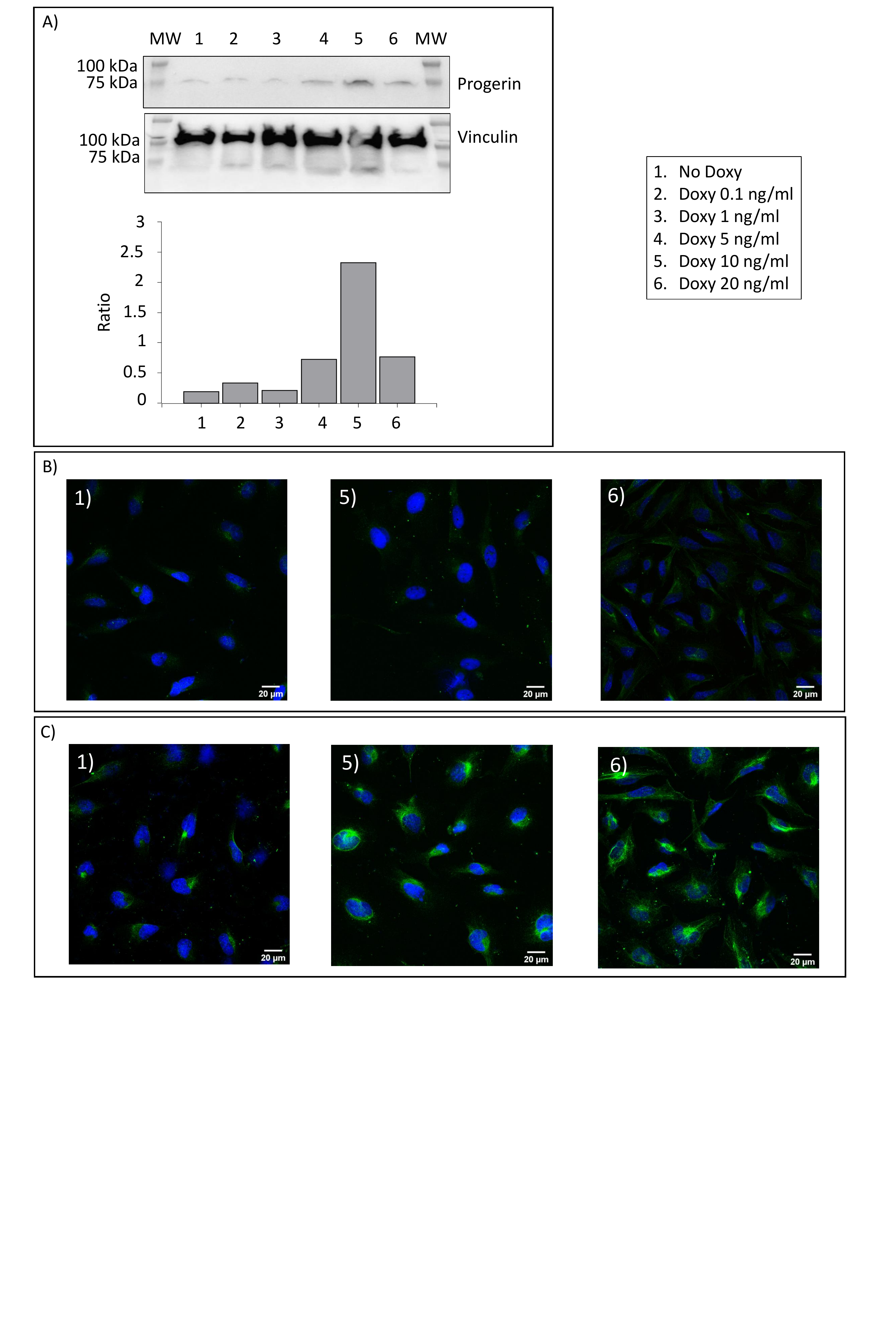}
\end{center}
\caption{\label{Fig:S1}
A)This panel reports a typical western blot of progerin in HeLa Tet-On 3G progerin-expressing cells exposed to increasing concentration of Doxy (0.1-20ng/ml). 20$\mu$g total protein were loaded on 10\% polyacrylamide gel, transferred on PVDF and incubated with anti-progerin (1:1000, ab66587, Abcam) overnight at 4$^\circ$C. Mouse anti-vinculin antibody (1:10000, V9264, Sigma) for 1 h at room temperature was used as housekeeping. Densitometric analysis was performed using ImageJ software. Y axis reports the ratio between the densitometric value of progerin with respect to the housekeeping (Ratio). B) Subconfluent HeLa Tet-On 3G ells were transfected with empty vector and fixed without treatment (1) or after the treatment with 10 ng/ml (5) and 20 ng/ml Doxy (6) with ice-cold methanol (see Materials and Methods for more details). The cells were incubated with anti-progerin (1:1000, ab66587, Abcam) at 4$^\circ$C overnight and then with AlexaFluor 488 (1:250, ab150113, Abcam) for 1h, at RT. Nuclei were stained with DAPI. Images were acquired by Leica SP2 laser scanning confocal microscope. Magnification is reported. 
C) Subconfluent HeLa Tet-On 3G cells transfected with $\Delta$50 lamin A vector
without Doxy treatment  (1) or after 10ng/ml (5) or 20ng/ml Doxy (6) Doxy treatment were fixed  with ice-cold methanol and incubated with anti-progerin (1:1000, ab66587, Abcam) at 4$^\circ$C overnight and then with AlexaFluor 488 (1:250, Abcam ab150113) for 1h, at RT. Nuclei were stained with DAPI. Images were acquired by Leica SP2 laser scanning confocal microscope. Magnification is reported. Data have been
collected over at least three independent experiments. }
\end{figure*}

\begin{figure*}[h]
\begin{center}
\includegraphics[width=12cm]{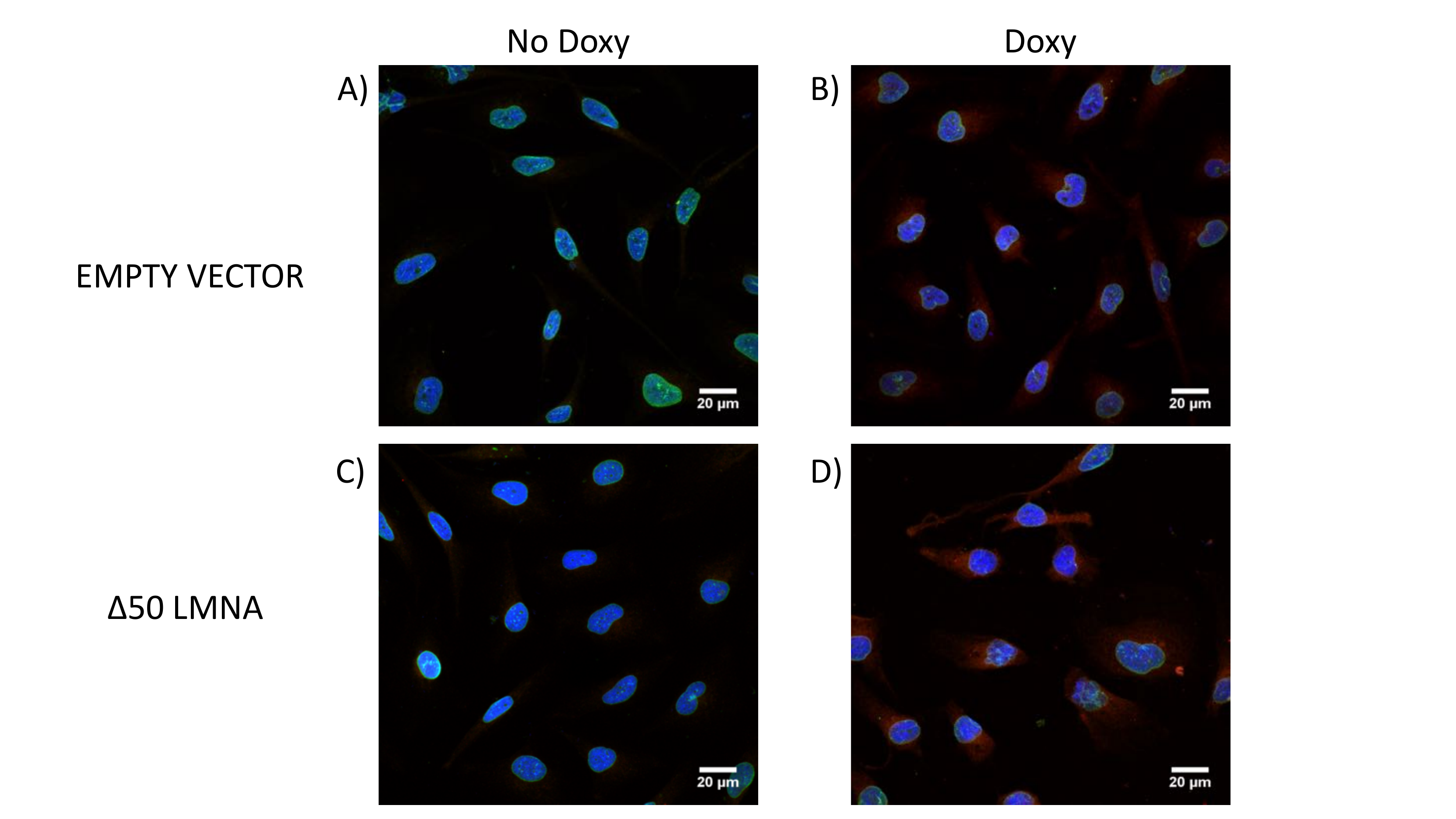}
\end{center}
\caption{\label{Fig:S2_new}
HeLa Tet-On containing empty vector and Hela Tet-On progerin expressing cells without Doxy tratment (No Doxy) or after induction with 10 ng/ml with Doxycycline (Doxy). Subconfluent cells were fixed with ice-cold methanol, then incubated with anti-PanLamin (1:50, ab20740, Abcam) at 4°C overnight and with AlexaFluo488 (1:250, ab15113, Abcam) for 1h, at RT. In red mCherry expression. Nuclei were stained with DAPI. Images were acquired by Leica SP2 laser scanning confocal microscope. Data have been
collected over at least three independent experiments. }
\end{figure*}

\begin{figure*}
\begin{center}
\includegraphics[width=12cm]{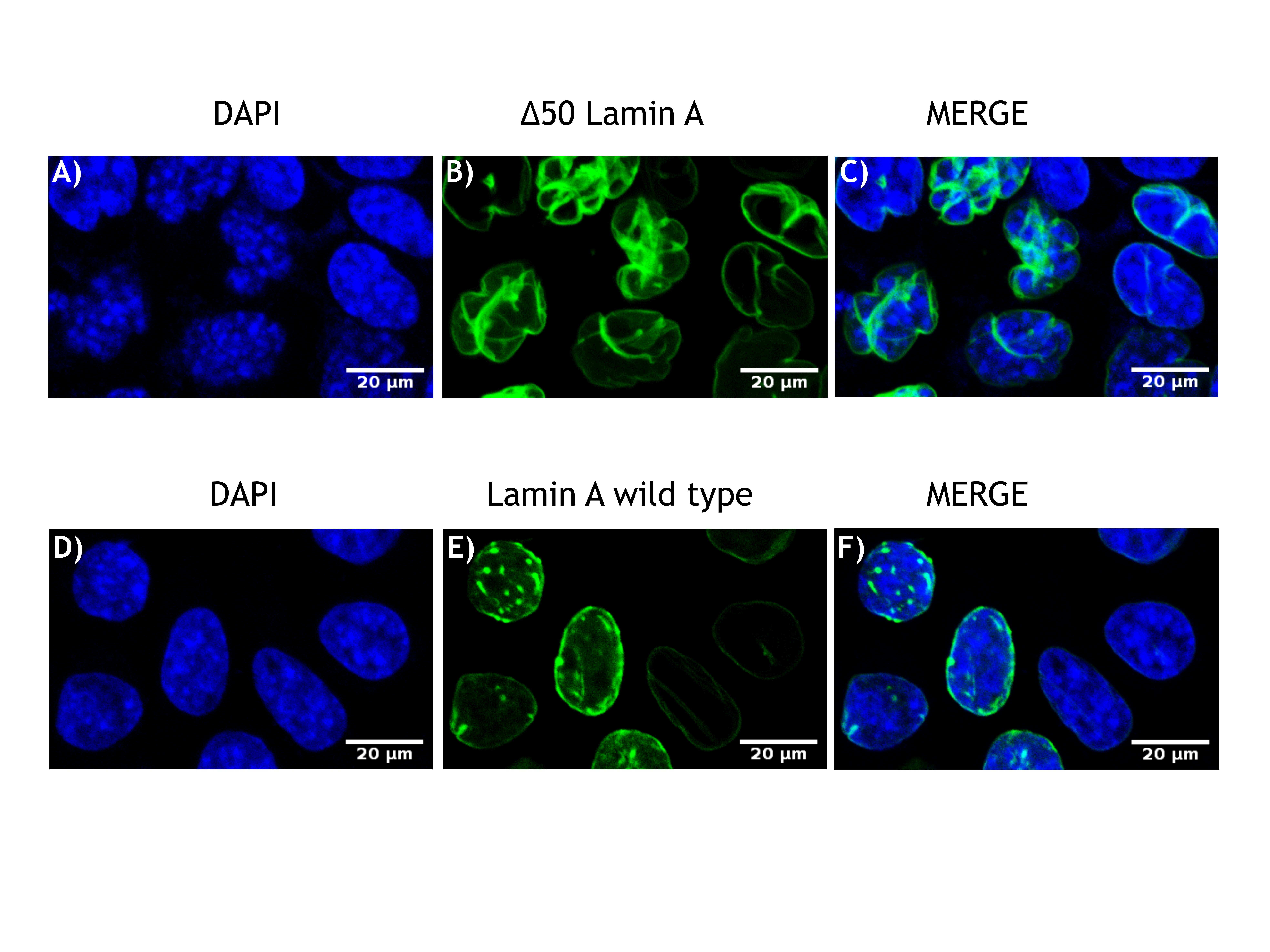}
\caption{\label{fig:emerald} Images of HeLa cell expressing mEmerald - wild type Lamin A (panel A-C) or pEGFP-$\Delta$50 LaminA ( panel  D-F), 48 hrs post-transfection.}
\end{center}
\end{figure*}

\begin{figure*}
\begin{center}
\includegraphics[width=12cm]{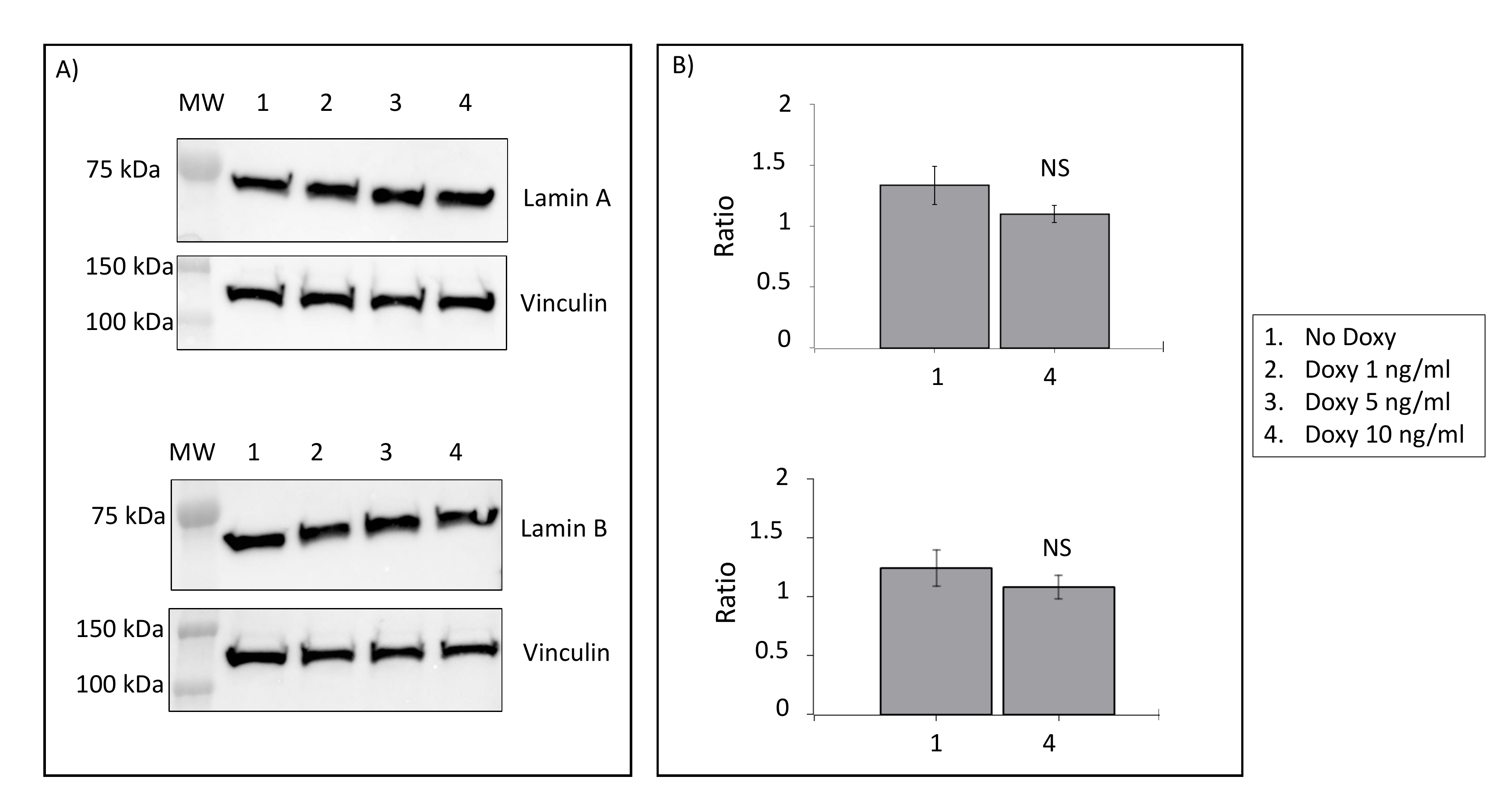}
\end{center}
\caption{\label{fig:S2}
A) A typical experiment showing of the level of expression of lamin A and B in HeLa Tet-On progerin expressing cells. 20$\mu$g total protein were loaded on 10\% polyacrylamide gel, transferred on PVDF and incubated with anti-lamin A (1:100, ab8980, Abcam) or anti-lamin B (1:200, ab16048, Abcam) overnight at $4^\circ$C. Anti-vinculin antibody (1:10000, V9264, Sigma) for 1 h at room temperature was used as housekeeping. 
B) Densitometric analysis of two independent experiments of western blot carried out as described in panel A. The Y axis reports the ratio between the mean densitometric value of lamin A or lamin B with respect to the corresponding housekeeping. Densitometric analysis was carried out using ImageJ software. Statistical significance was established by the 
{\it t-test}. NS: not significant. }
\end{figure*}

\begin{figure*}
\begin{center}
\includegraphics[width=12cm]{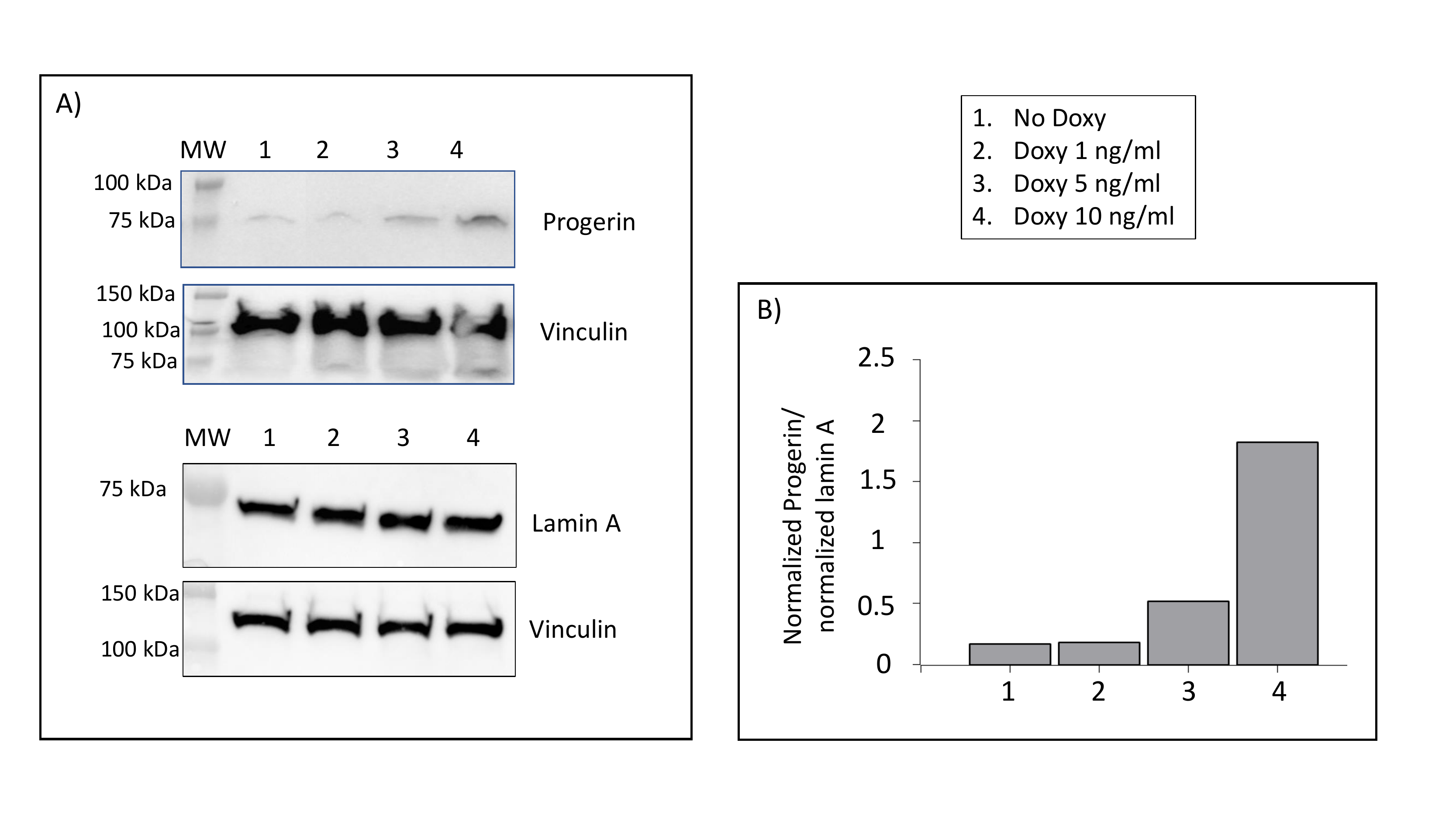}
\end{center}
\caption{\label{fig:S3}
Typical experiment of the ratio between the level of expression of progerin and lamin A in HeLa Tet-On progerin expressing cells.  20$\mu$g total protein of HeLa Tet-On progerin expressing cells were loaded on 10\% polyacrylamide gel, transferred on PVDF and incubated with anti-progerin (1:1000,  ab66587, Abcam) or anti-lamin A (1:100, cod.ab8980, Abcam) overnight at 4$^\circ$C. Mouse anti-vinculin antibody (1:10000, V9264, Sigma) for 1h at room temperature was used as housekeeping. The Y axis (normalized progerin/normalized lamin A) represents the ratio between the  denistometric value of progerin with respect to  lamin A both normalized with respect to the corresponding denistometric values of the housekeeping. The experiment shown here is the same reported in Fig. \ref{Fig:S1} (progerin) and Fig. \ref{fig:S2} (lamin A).}
\end{figure*}

\begin{figure*}
\begin{center}
\includegraphics[width=15cm]{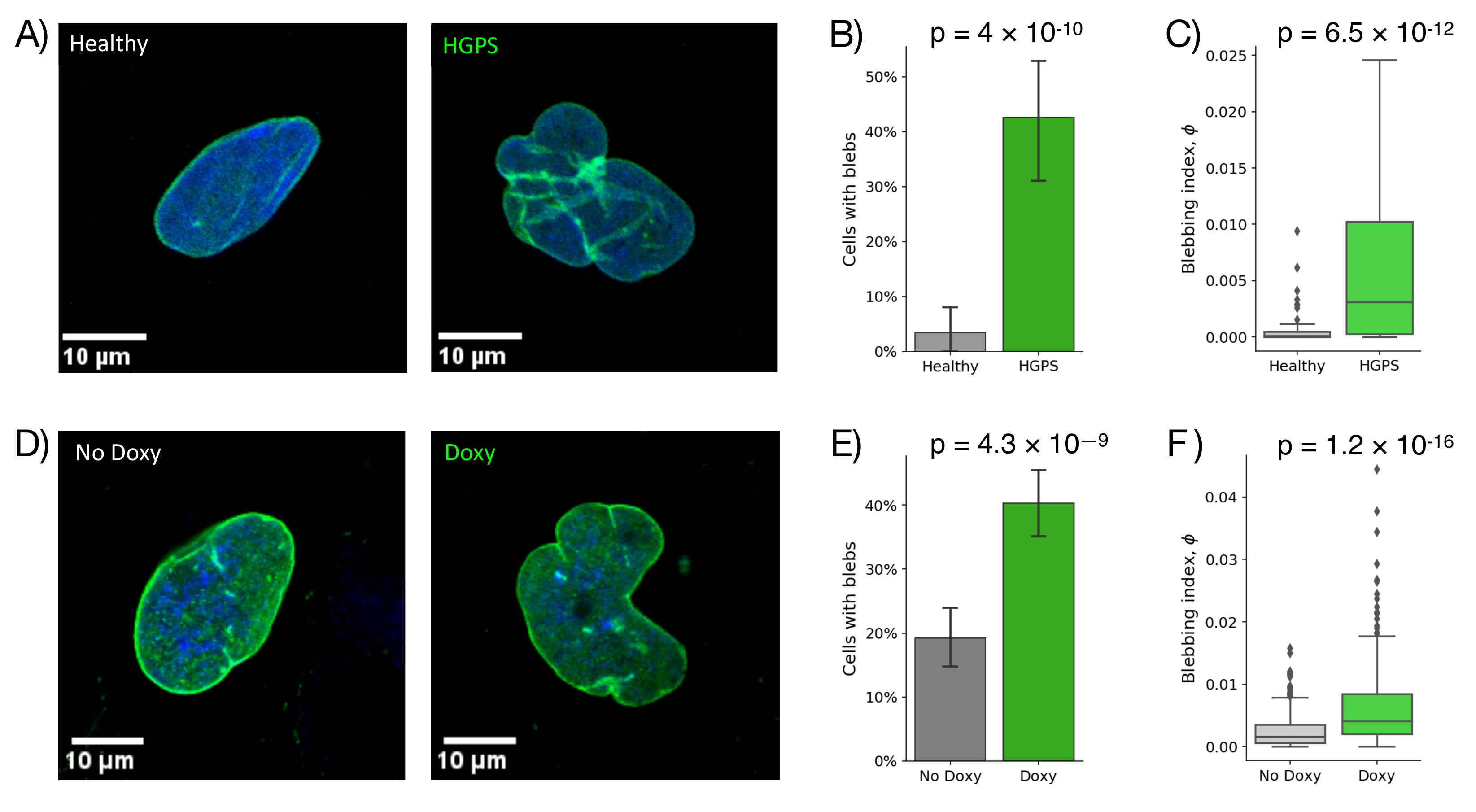}
\end{center}
\caption{\label{fig:S4}
Progerin expression affects nuclear morphology.  Typical nuclei obtained from subconfluent A) HGPS cells (HGPS), healthy mother of HGPS patient  (Healthy), or D) HeLa Tet-On cells without Doxy tratment (No Doxy) or after induction with 10 ng/ml with Doxy (Doxy). Subconfluent cells were fixed with ice-cold methanol, then incubated anti-PanLamin (1:50 ab20740, Abcam) at 4$^\circ$C overnight and with AlexaFluo488 (1:250, ab15113, Abcam) for 1h, at RT. Nuclei were stained with DAPI. Images were acquired by Leica SP2 laser scanning confocal microscope. Quantification of morphological alterations of (B-C) HGPS cells, mother of HGPS cells, (E-F) HeLa Tet ON progerin expressing cells (Doxy) and HeLa Tet-On cells without Doxy treatment (No Doxy) is performed by computing 
(B and E) the number of cells with blebs and (C and F) blebbing index as described in the material and method section. Statistics was performed over 297 No Doxy nuclei, 330 Doxy nuclei, 87 healthy nuclei, and 87 HGPS nuclei. The $p$-values reported are obtained via a Kolmogorov-Smirnov test comparing the distribution of the relevant magnitude among the two groups (Doxy vs No Doxy or Healthy vs HGPS). The computation is implemented via ``ks\_2samp'' function from SciPy library \cite{SciPy}. Data have been
collected over at least three independent experiments.  }.
\end{figure*}

\begin{figure*}
\begin{center}
\includegraphics[width=15cm]{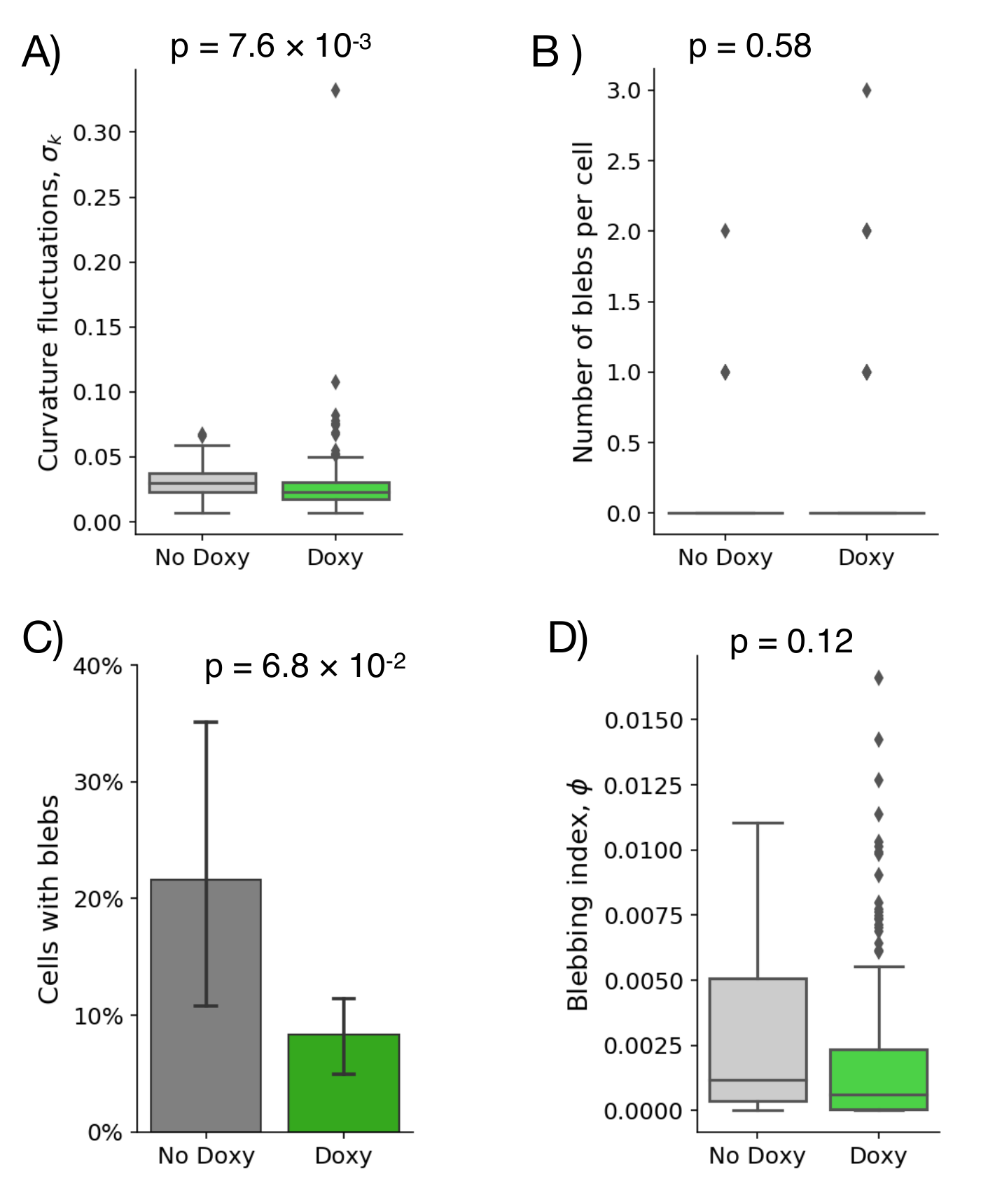}
\end{center}
\caption{\label{fig:empty} When HeLa cells are transfected with an empty vetcor, Doxy treatment
has no significant effect on nuclear morphology. The panels report A) curvature fluctuations, B) number of blebs per cell, C) cells with blebs, D) blebbiness index. Data have been
collected over at least three independent experiments. 
}
\end{figure*}

\begin{figure*}
\begin{center}
\includegraphics[width=12cm]{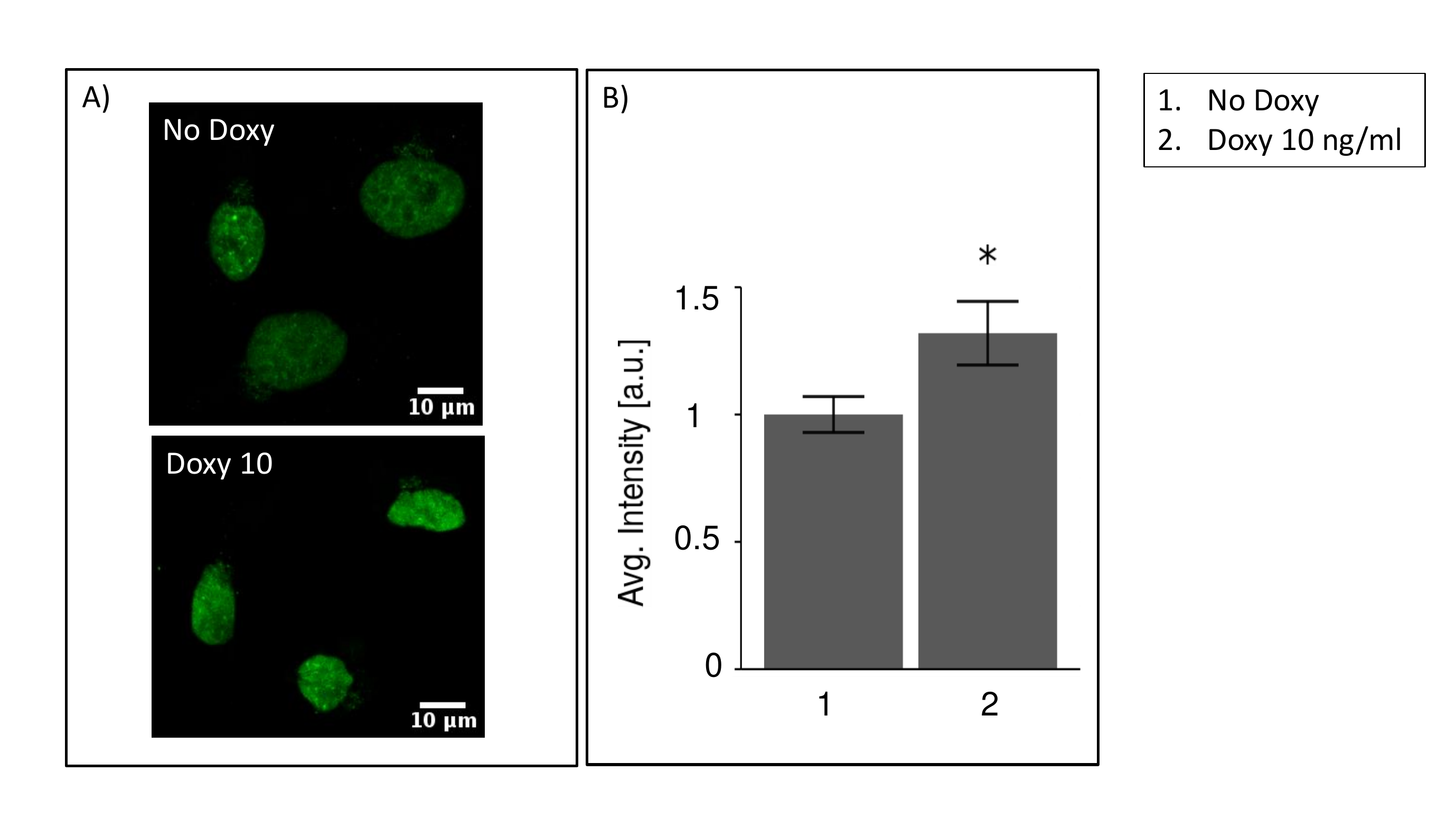}
\end{center}
\caption{\label{fig:S5}
Immunofluorescence analysis of HeLa Tet-On progerin expressing cells with respect to non expressing one (No Doxy). Subconfluent cells were fixed with ice cold 100\% methanol for 5min at -20$^\circ$C, incubated with anti-HP1 (1:250, ab109028, Abcam) overnight at $4^\circ$C and with the secondary antibody for 1 h at RT (for more details see Materials and Method section). The nuclei were counterstained with DAPI and the slides mounted with Pro-long anti fade reagent (cod P36931, Life technologies). A) The images are acquired with a Leica TCS NT confocal microscope. Immunofluorescence intensity was estimated using customized ImageJ macro evaluating  single pixel fluorescence after subtracting the background noise. B) The average fluorescence was calculated on pixels that passed the background filtering. In all the analyzed frames, nuclei close to the edge of the image or superimposed were manually discarded. Plot shows the average HP1 fluorescence for HeLa Tet-On progerin expressing cells (8 replica for a total of 66 cells) and without  Doxy induction (6 replica,  for a total of 45 cells) . Error bars indicate standard errors (SE). Statistical significance was assessed using unpaired t-test (* p<0.05).}
\end{figure*}

\begin{figure*}
\begin{center}
\includegraphics[width=12cm]{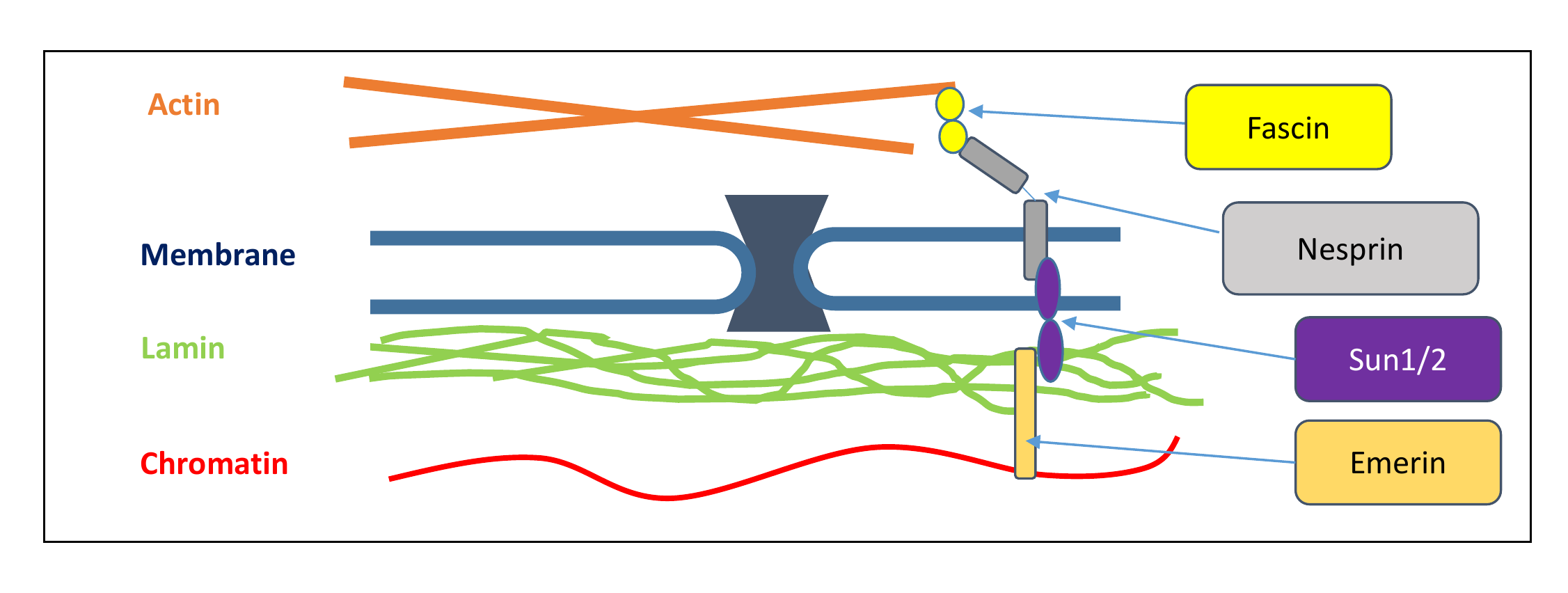}
\caption{\label{fig:S6} Schematic picture of the main tethering factors linking cytoskeletal F-actin, lamin shell and chromatin.}
\end{center}

\end{figure*}

\begin{figure*}
\begin{center}
\includegraphics[width=12cm]{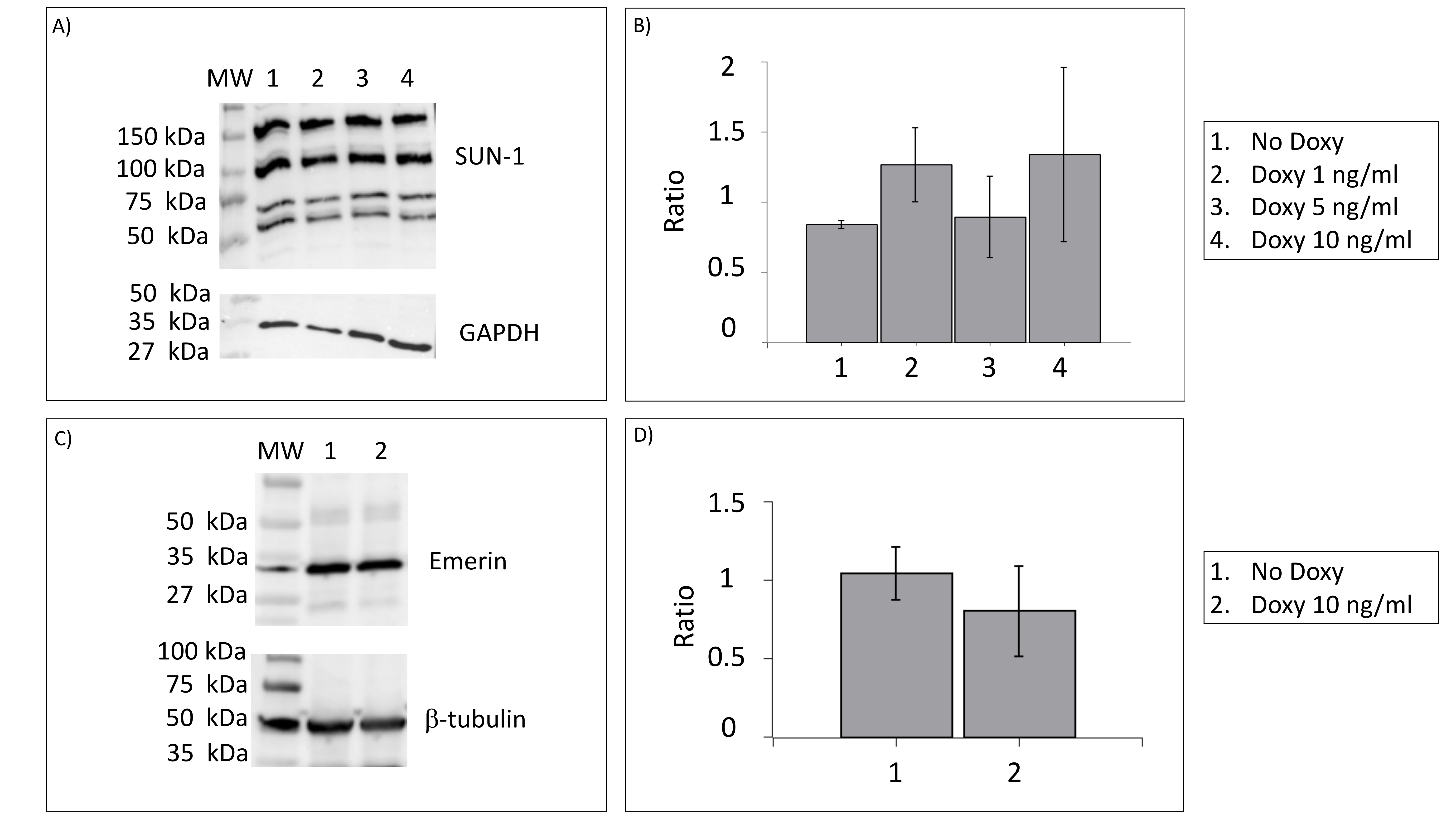}
\caption{\label{fig:S7} A and C) Typical western blot of SUN1 and emerin in HeLa Tet-On progerin expressing cells without the induction with Doxy and after Doxy (from 1 to 10ng/ml). 20$\mu$g total protein were loaded on 10\% polyacrylamide gel, transferred on PVDF and incubated with rabbit anti-SUN1 (1:1000, ab103021, Abcam) or anti-emerin (1:200, ab40699, Abcam) overnight at 4$^\circ$C. GAPDH (1:5000 G9545, Sigma) or $\beta$-tubulin (1:5000, T8328, Sigma) are used as housekeeping. Molecular weight is indicated by MW. B and D) Mean of densitometric analysis of two independent western blot of SUN1 and emerin performed with ImageJ. The Y axis shows the ratio between the densitometric value of B) SUN1 or D) emerin with respect to the corresponding housekeeping.}
\end{center}
\end{figure*}

\begin{figure*}
\begin{center}
\includegraphics[width=12cm]{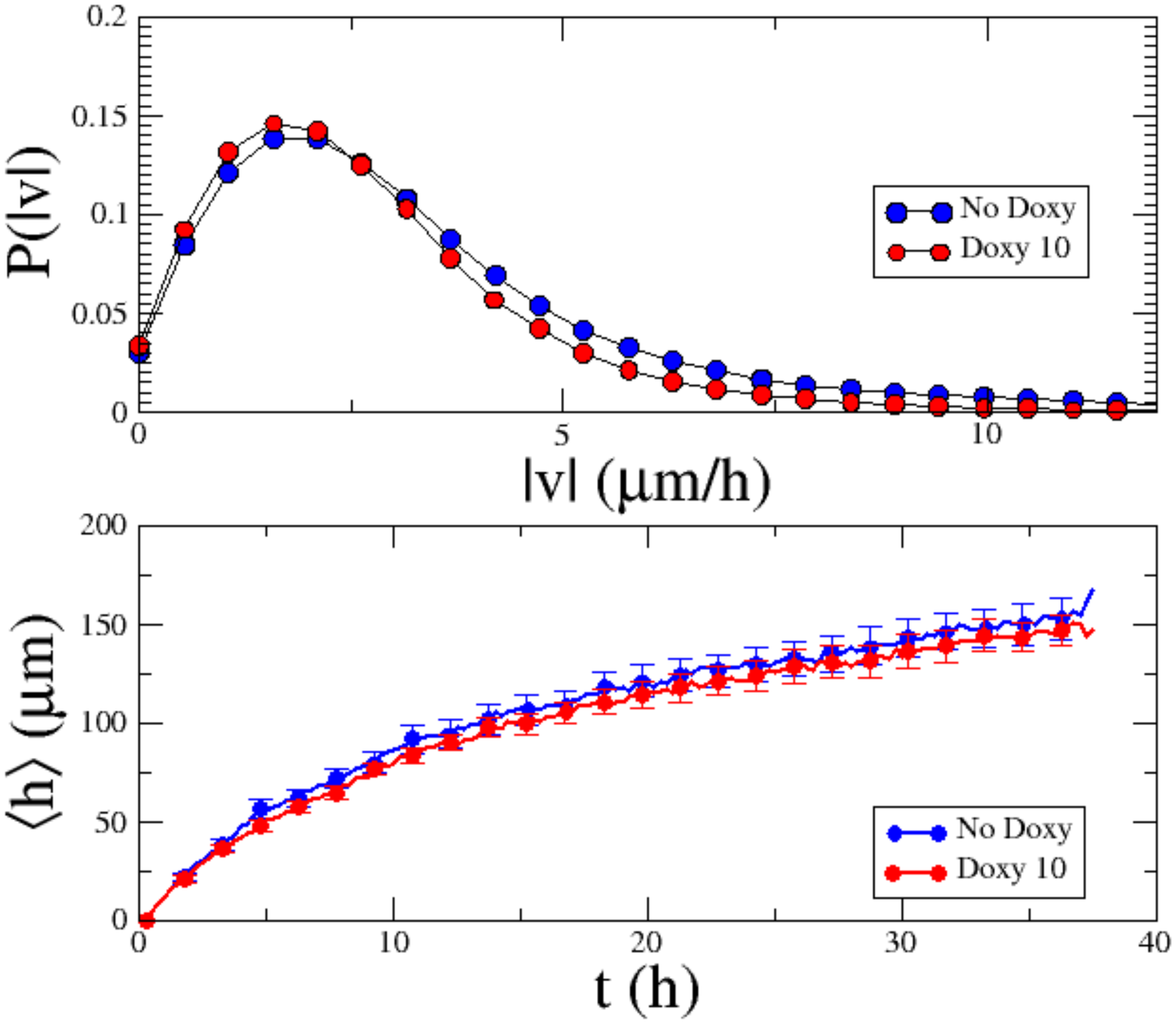}
\caption{\label{fig:S8}
A) The velocity distributions obtained from PIV in scratch assays in HeLa TetON progerin-expressing cells after 10 ng/ml Doxy treatment and in control untreated cells. B) The time evolution of the average position of the front. The data are averaged over four independent experiments for each condition.}
\end{center}
\end{figure*}

 \begin{figure*}
\begin{center}
\includegraphics[width=12cm]{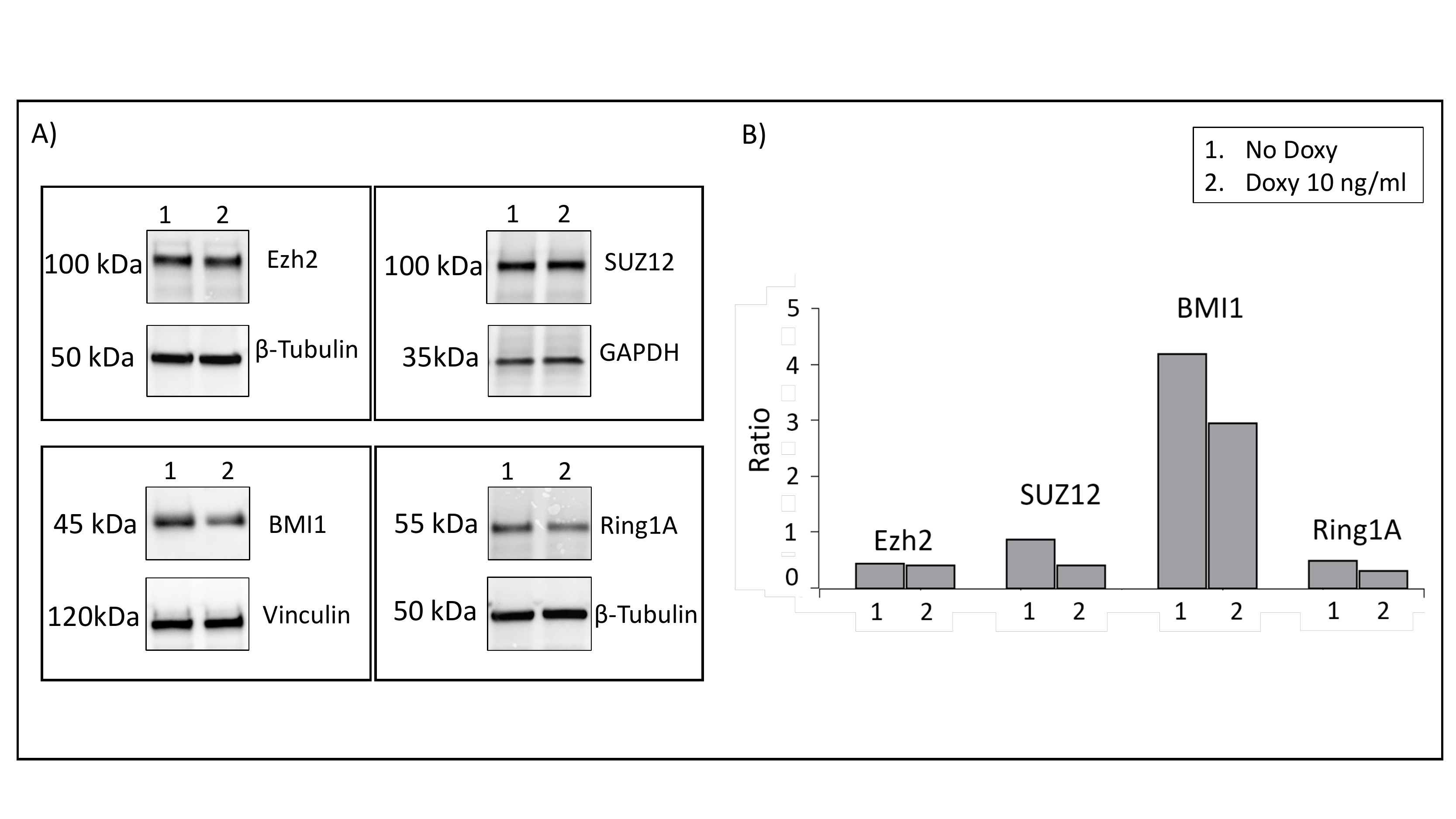}
\caption{\label{fig:S9} A) Typical western blot of PgC proteins in HeLa Tet-On progerin expressing cells without the induction with Doxy and after Doxy (10 ng/ml). 20 $\mu$g total protein were loaded on 10\% polyacrylamide gel, transferred on PVDF and incubated with anti SUZ12 (1:1000, 3737, Cell Signalling); anti-EZH2 (1:1000, 5246, Cell Signalling); anti Ring1a (1:1000, 13069, Cell Signalling); anti-BMI1 (1:1000, 6964, Cell Signalling) overnight at 4$^\circ$C. Anti-vinculin (1:10000, V9264, Sigma) or anti-GAPDH (1:5000, G9545, Sigma) or $\beta$-tubulin (1:5000, T8328, Sigma) for 1 h at room temperature was used as housekeeping. Panel B) Densitometric analysis of the gel shown in Panel A by ImageJ software. The Y axis shows the ratio between the densitometric value of PcG proteins with respect to the corresponding housekeeping. }
\end{center}
\end{figure*}

\begin{figure*}
\begin{center}\textbf{\begin{flushleft}
•\emph{\textit{•}}
\end{flushleft}}
\includegraphics[width=12cm]{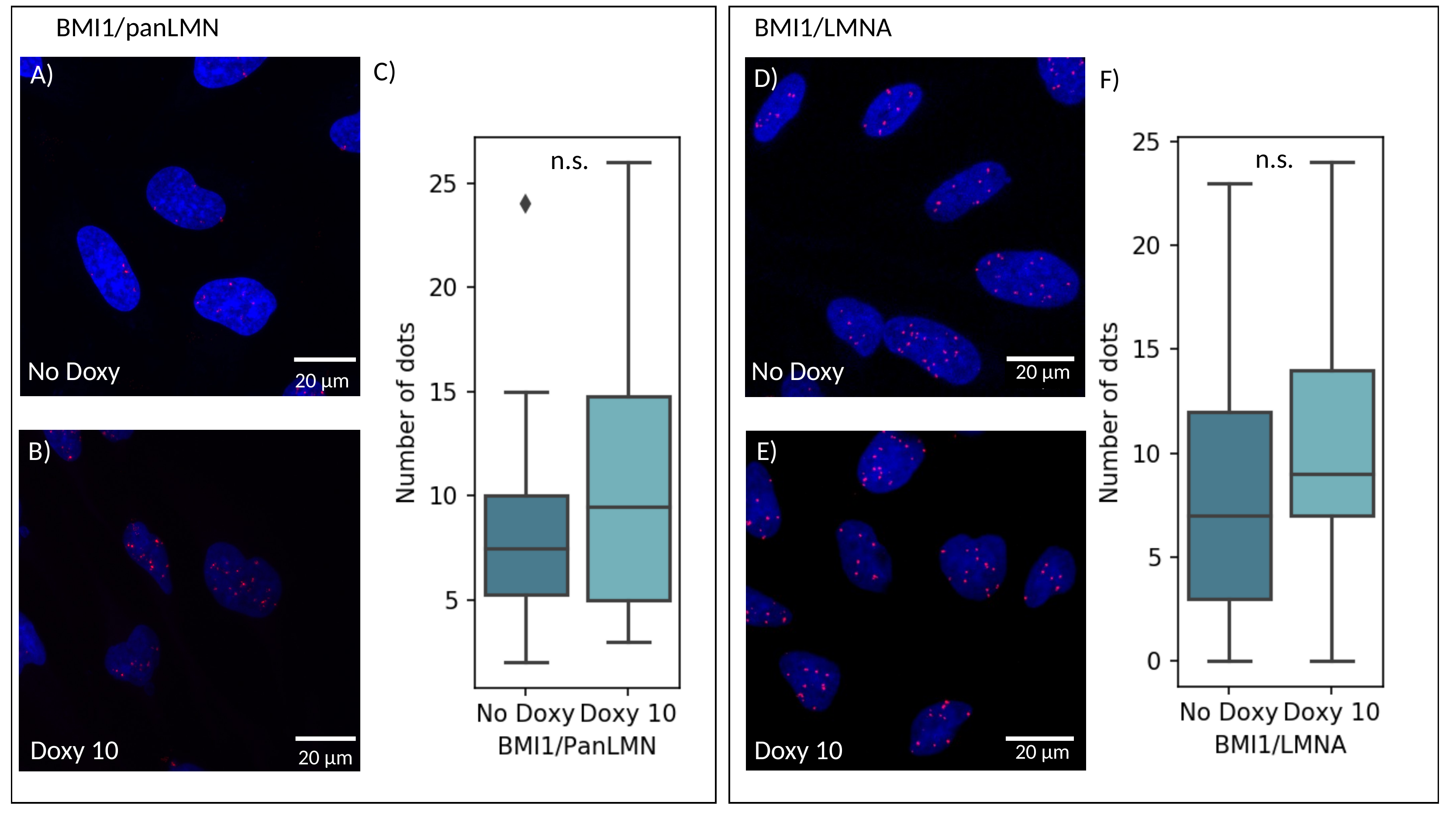}
\caption{\label{fig:S10}
The tethering is quantified by the proximity ligation assay measuring the interaction between all lamins (panLMN) or lamin A  and PcG BMI1. Briefly, subconfluent cells were fixed on slides with ice cold 100\% methanol for 5 min. Slides were then incubated  in a humidity chamber overnight at 4$^\circ$C with  PanLamin (1: 50, cod. 20740, Abcam) or anti-lamin A (1:100 cod. ab8980, Abcam)  antibody with  BMI1 (1:600, mAb 6964, Cell Signaling). After washing, samples were incubated in a pre-heated humidity chamber for 1 hour at 37$^\circ$C with anti-rabbit PLUS and anti-mouse MINUS PLA probes diluted 1:5. Ligation and amplification steps were performed according to manufacturer's instructions.  Slides were mounted with Duolink In Situ Medium. Panel A), B), D) and E)  show typical experiments for each experimental conditions. The number of aggregates linking BMI1 with C) Panlamin or F) lamin A are quantified as described in the Materials and Methods section.  The analysis was performed on 72  nuclei (42 without Doxy and 30 with Doxy) for PanLamin and 134 nuclei (81 without 53 with Doxy) for lamin A. Data have been
collected over at least three independent experiments. }
\end{center}
\end{figure*}

\begin{figure*}
\begin{center}
\includegraphics[width=12cm]{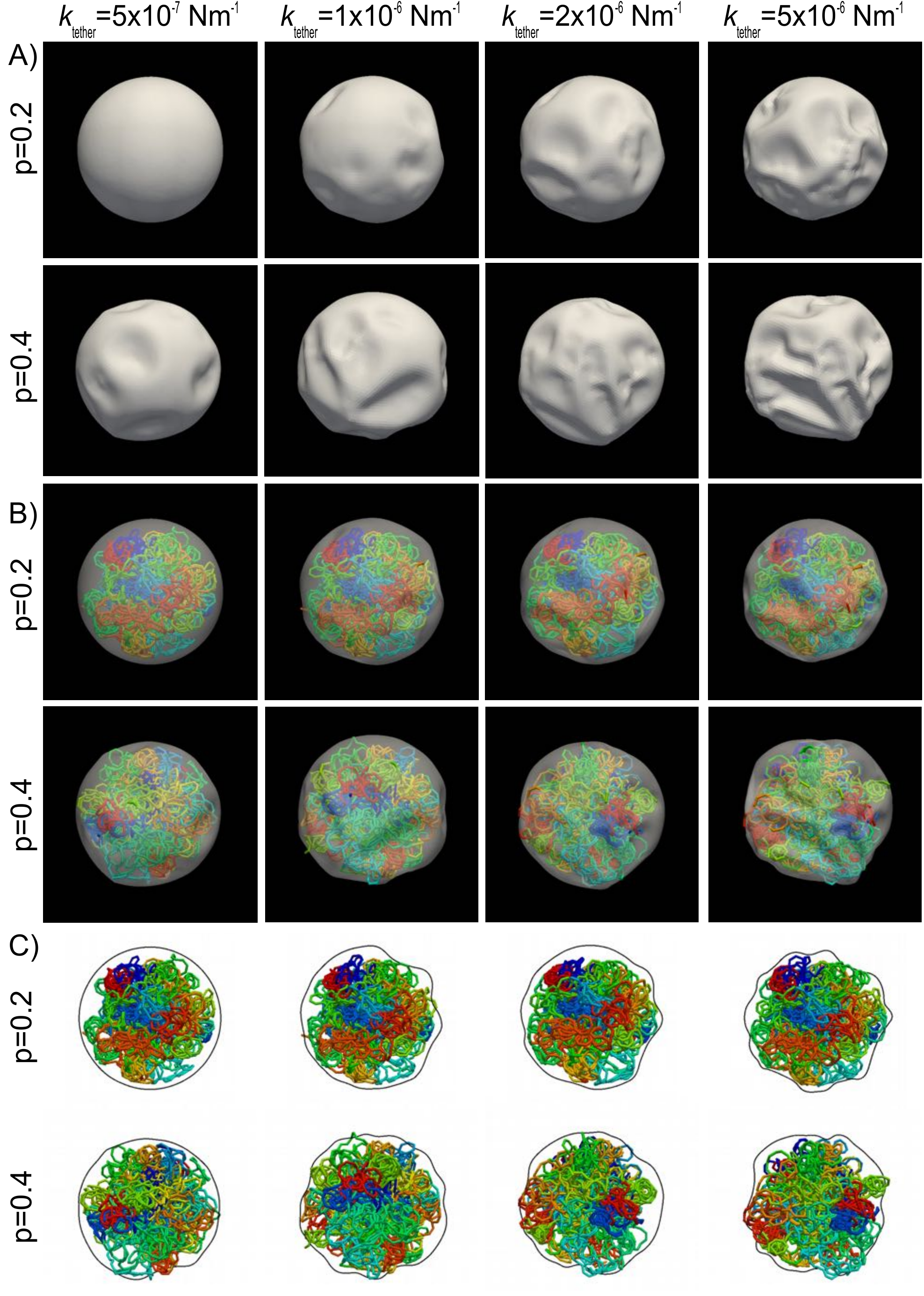}
\end{center}
\caption{Without lamin domains, blebs do not form. We show here a range of tethering strengths and probabilities; in all cases the tethers are distributed uniformly at random on the nuclear envelope. Where a higher tether density occurs by chance, the envelope is pulled inwards, but no corresponding outward deformation (bleb) occurs.}
\label{fig:random_noblebs}
\end{figure*}

\begin{figure}
\begin{center}
\includegraphics[width=12cm]{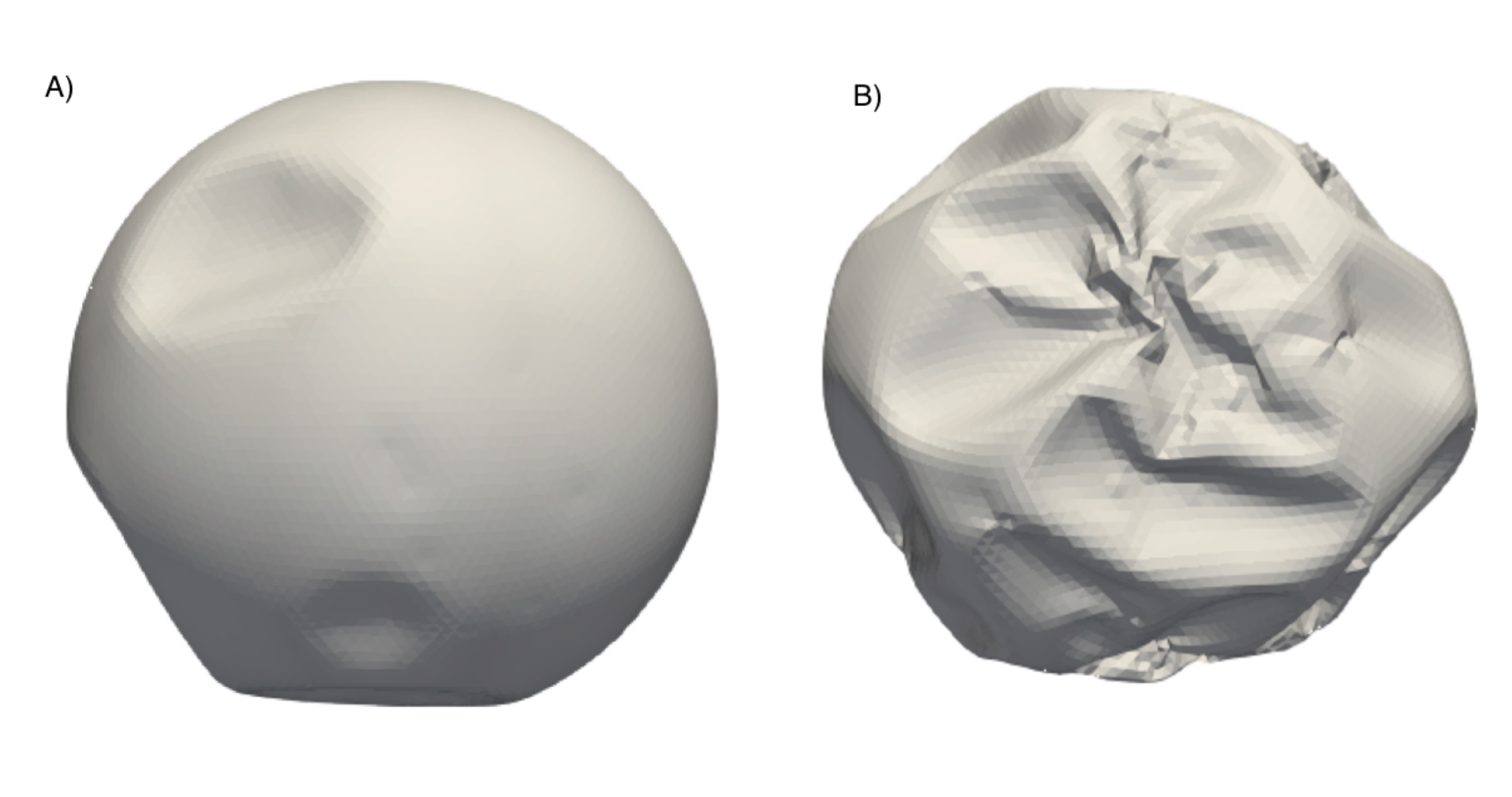}
\end{center}
\caption{Without lamin domains, blebs do not form, even when chromatin-nuclear envelope tethering is non-uniformly distributed. Panel a shows the nuclear envelope configuration when tethers are concentrated in a single octant of the sphere. Panel b shows the case where tethers are concentrated along a line. In both cases, high tether densities make the nuclear envelope fold inwards rather than bleb outwards.}
\label{fig:regions_lines}
\end{figure}

\begin{figure}
\begin{center}
\includegraphics[width=12cm]{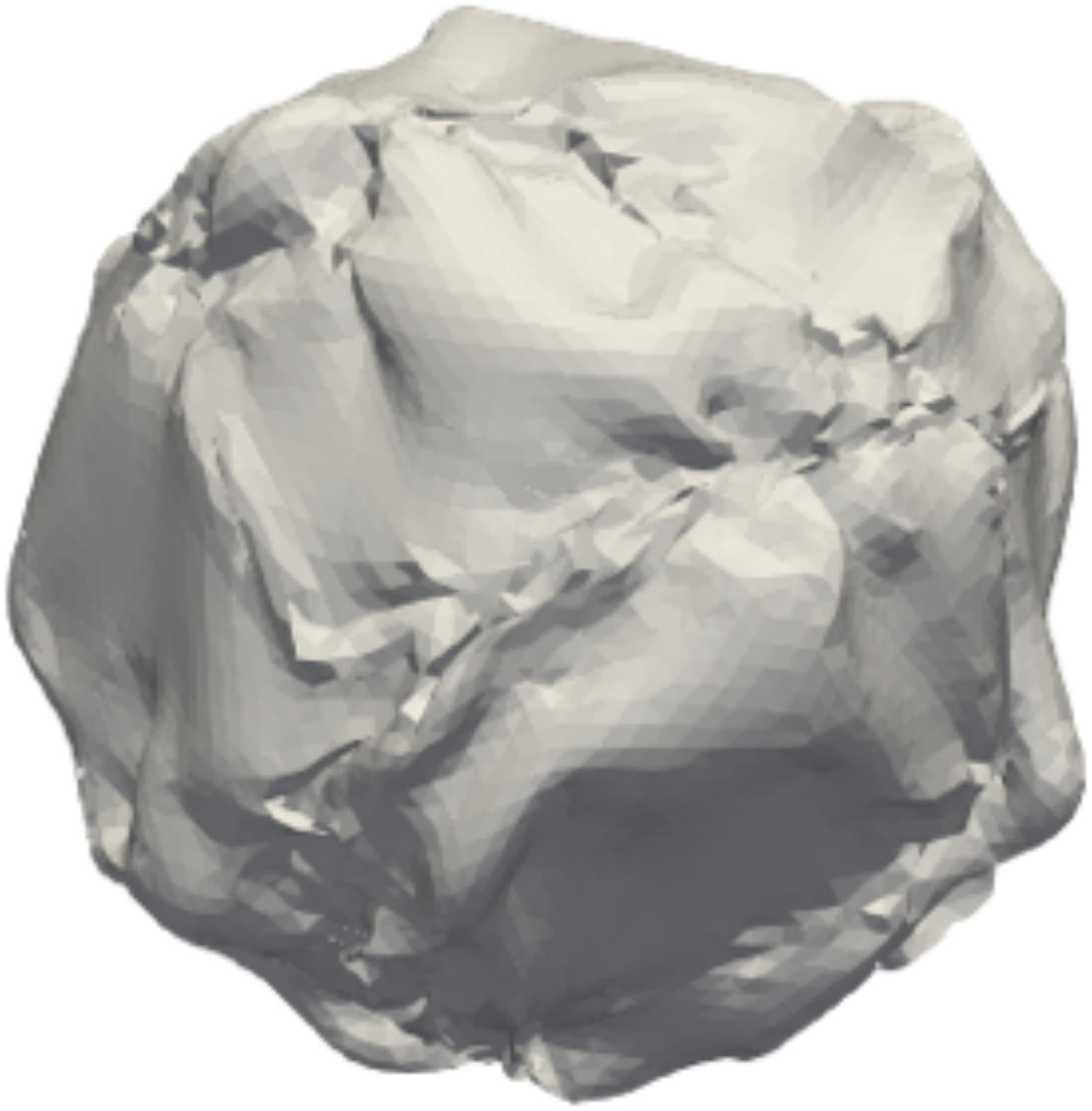}
\end{center}
\caption{Smooth deformation of the nuclear envelope only occurs if its bending stiffness is sufficiently large. Here the spring constant giving bending stiffness is $10^{-16}$ J rad$^{-2}$  inside the lamin domains and $10^{-18}$ J rad$^{-2}$ on the domain walls, resulting in a crumpled nuclear envelope.}
\label{fig:crumple}
\end{figure}

\begin{figure}
\begin{center}
\includegraphics[width=12cm]{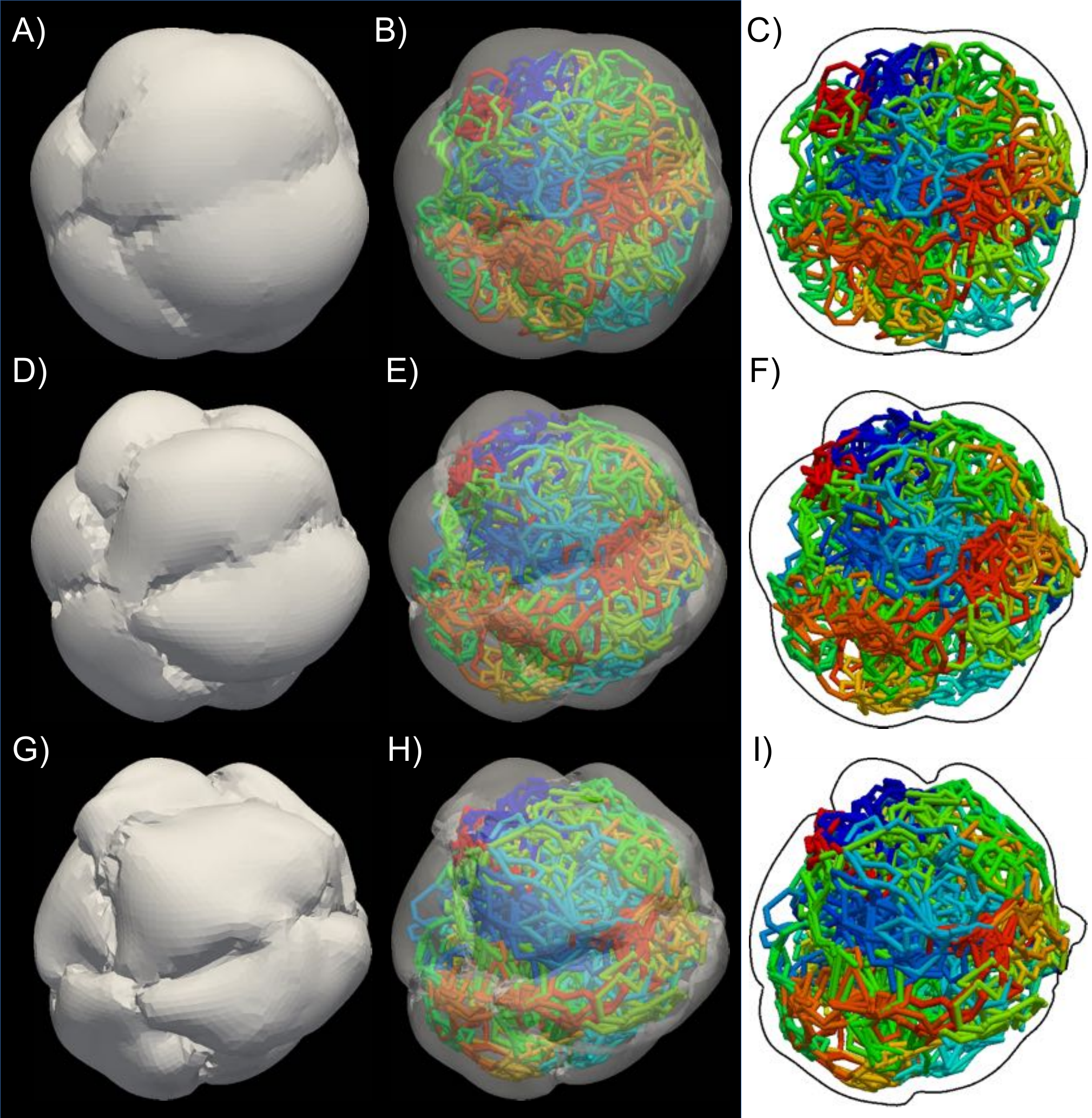}
\end{center}
\caption{Role of the chromatin tether elasticity in nuclear morphology. 
Tethering strength has value A-B-C)10$^{-6}$ Nm$^{-1}$, D-E-F)10$^{-5}$ Nm$^{-1}$ and 
G-H-I) 10$^{-4}$ Nm$^{-1}$. Smaller tethering strength leads to less pronounced blebs.
Panels in the second columns show the same configurations with the lamin shell made transparent in order to visualize the chromatin inside. Panels in third columns show instead a slice taken of the nuclear envelope for the same configuration. In all cases, the link density is $p=0.5$.}
\label{fig:S14}
\end{figure}

\clearpage

\begin{table*}[h]
\begin{center}
\includegraphics[width=12cm]{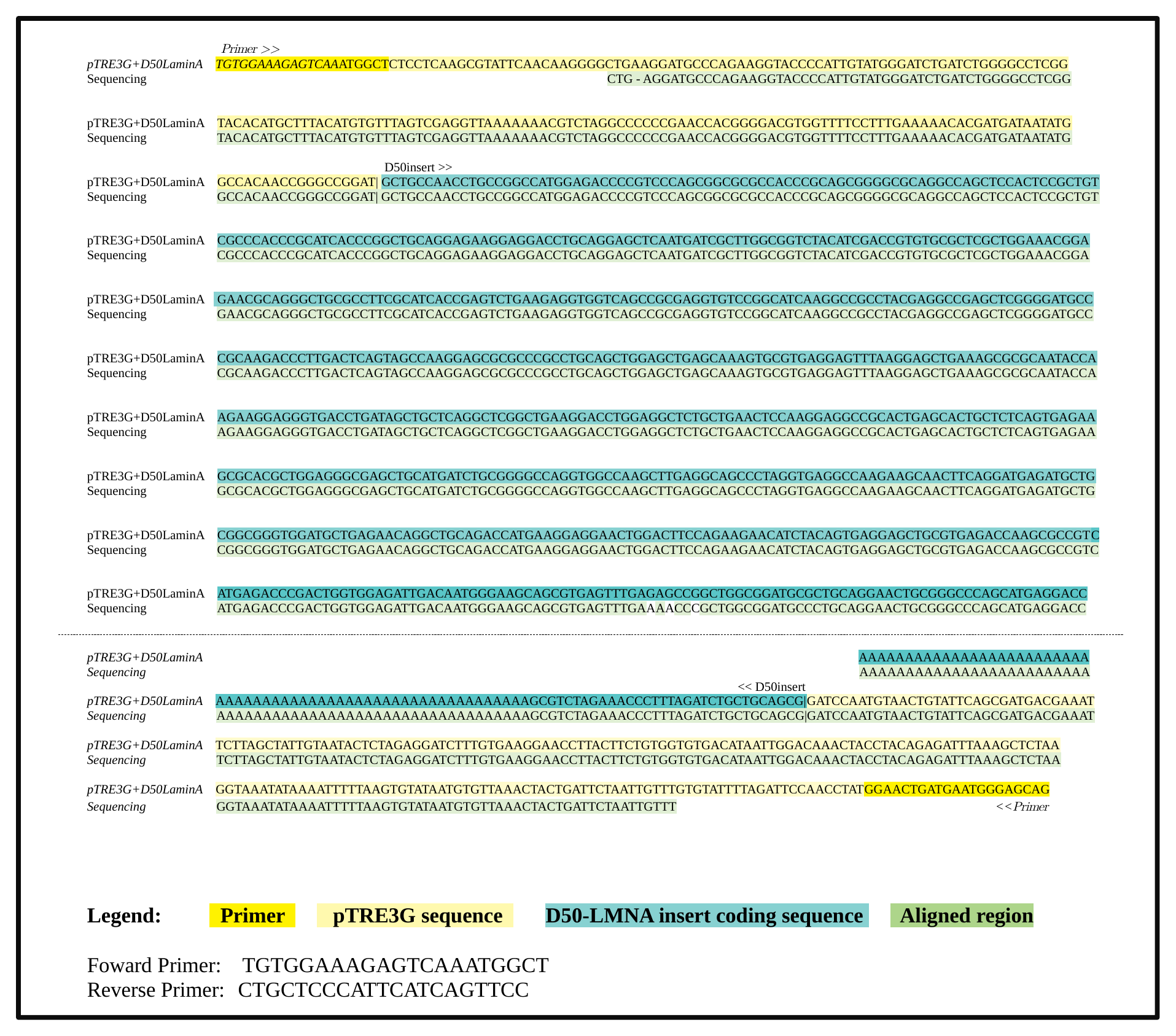}
\end{center}
\caption{\label{TableS1} $\Delta$50-lamin A sequence obtained from pEGFP-D50 lamin A plasmid (Addgene, 17653) was inserted in pTRE3G-mCherry vector (Clontech, 631165) as described in Materials and Methods section. In this figure is shown the alignment between expected sequence of the final plasmid with $\Delta$50 insert and the result of clone sequencing obtained using T-Coffee\cite{Tcoffee}. IRES forward sequencing primer and pTRE3G-mCherry vector as well as the region of $\Delta$50-lamin A that align to LMNA/C gene are reported (yellow and light blue as in legend). Start of D50-insertion coincide with AfeI/EcorV excision point. Sequencing confirm the presence of all the coding part of $\Delta$50-lamin A insert at expected position (green region). }
\end{table*}

\begin{table}[!h]
\begin{center}
\begin{tabular}{ |c|c|c| } 
\hline
Biological element & Number of beads & Organization \\
\hline
\hline
Nuclear lamina & 10242 & initialized as a sphere of radius 6 $\mu$m\\
\hline
Chromatin & 5888 & 46 polymers with 128 monomers each \\
\hline 
Cytoskeleton & 100 & randomly distributed around the nuclear lamina \\
\hline
\end{tabular}
\begin{tabular}{ |c|c|c|c| } 
\hline
\hline
Bond link & $K$ [10$^{-3}$N/m]  & $r_0$ [10$^{-6}$m] & \\
\hline
\hline
lamina beads &  5.0  &  0.15 - 0.25 &  \\ 
\hline
monomers of chromatin filament & 1.0 & 0.6 & \\
\hline
lamina - chromatin ($k_{tether}$ with density $\rho$) & 0.1  & 0.1 &  \\ 
\hline
lamina - cytoskeleton ($k_{cyto}$)  & 1,5,10 & 0.8 - 12 & \\
\hline
\hline
Improper & $k$ [10$^{-15}$J rad$^{-2}$]  & $\chi_0$ [deg] & \\
\hline
\hline
lamina beads &  1.0  &  180  &\\  
\hline
lamina beads - domain walls & 0.01 & 180  &\\ 
\hline
\hline
Harmonic angle & $\kappa$ [10$^{-15}$J rad$^{-2}$]  & $\theta_0$ [deg] & \\
\hline
\hline
monomers of chromatin filament &  0.2  &  130 &  \\  
\hline
\hline
LJ interaction & $\epsilon$ [10$^{-15}$J]  & $\sigma$ [10$^{-6}$m] & $r_c$ [10$^{-6}$m]  \\
\hline
\hline
chromatin - chromatin & 1.0  & 0.12   & 0.2\\
\hline
lamina - chromatin & 1.0 & 0.6 & 0.6 \\
\hline
\end{tabular}
\captionof{table}{A summary of the parameters of the simulation model. The harmonic bond potential defined as $E=K/2(r-r_0)^2$ where $r_0$ is the equilibrium bond distance between two beads. Harmonic impropers are defined over 4 beads with an energy $E=k/2(\chi -\chi_0)^2$
where $\chi$ is the improper angle, $\chi_0$ the equilibrium 
value and $k$ is the surface bending stiffness. Harmonic angles with energy
$E=\kappa/2(\theta -\theta_0)^2$ being $\theta_0$ the equilibrium value of the angle and $\kappa$ 
is line bending stiffness. Standard 12/6 Lennard-Jones potential with the parameters for the energy~$\epsilon$, the zero-crossing distance for the potential~$\sigma$, 
and cutoff radius~$r_c$ 
}\label{table:parameters}
\end{center}
\end{table}

\clearpage

\section*{Supplementary video caption}
{\bf Video S1.} A simulation of the nuclear fluctuations induced by cytoskeletal contractions.
Local expansion is marked in red and contraction in blue.

\end{document}